\documentclass[preprint,12pt]{elsarticle}

\usepackage{hyperref}

\makeatletter
\providecommand{\doi}[1]{%
  \begingroup
    \let\bibinfo\@secondoftwo
    \urlstyle{rm}%
    \href{http://dx.doi.org/#1}{%
      doi:\discretionary{}{}{}%
      \nolinkurl{#1}%
    }%
  \endgroup
}
\makeatother

\biboptions{numbers,sort&compress}

\usepackage{etoolbox}
\makeatletter
\patchcmd{\ps@pprintTitle}{\footnotesize\itshape
       Preprint submitted to \ifx\@journal\@empty Elsevier
       \else\@journal\fi\hfill\today}{\relax}{}{}
\makeatother

\usepackage{amssymb,graphicx,subfigure}
\usepackage[margin=3cm]{geometry} 
\usepackage{amsmath}
\usepackage{color}

\usepackage{lipsum}
\usepackage{setspace,url}

\newcommand{\rmr}{\mathrm{r}}  
\newcommand{\rmz}{\mathrm{z}}  
\begin{document}

\begin{frontmatter}
 
\title{Modelling hollow thermoplastic syntactic foams under high-strain compressive loading}
 
\author[label1]{Michael J.~A.~Smith\corref{cor1}}
\address[label1]{Department of Mathematics,  University of Manchester, Oxford Rd, Manchester M13 9PL, UK}
\ead{ms2663@cam.ac.uk}
\cortext[cor1]{Corresponding author}

\author[label1,label2]{Zeshan Yousaf\corref{cor2}}
\address[label2]{Robotics and Textile Composite Group, Northwest Composite Centre, Department of Materials,  \\ University of Manchester, Manchester M13 9PL, UK}
\author[label2]{Prasad Potluri\corref{cor2}}
\author[label1]{William J.~Parnell\corref{cor2}}

\begin{abstract}
The  mechanical response    of    syntactic foams  comprising hollow thermoplastic microspheres (HTMs) embedded in a polyurethane matrix were experimentally examined under uniaxial compressive strain. Phenomenological strain energy models  were subsequently  developed to capture both the axial stress-strain and transverse strain response of the foams.       HTM syntactic foams were found to  exhibit increased small-strain stiffness with  reduced density, revealing a  highly-tuneable   and extremely lightweight syntactic foam  blend for   applications.    The foams   were also found to become   strongly compressible at large strains and possess  a high threshold for plastic deformation, making them a   robust alternative to hollow glass microsphere    syntactic foams.  The non-standard transverse strain relationship exhibited by HTM syntactic foams at high filling fractions     was captured by      Ogden-type strain energy models. The thermal characteristics of these syntactic foams were also explored with Differential Scanning Calorimetry   testing which showed that HTMs have  a negligible impact on the thermal characteristics of the matrix.
\end{abstract}

 
\end{frontmatter}

\section{Introduction}
\label{sec:intro}
Syntactic foams are composite materials comprising a suspension of   gas-filled  microspheres (microballoons)  within a   matrix material \cite{gupta2014applications}. These foams are well-known for their enhanced mechanical performance,  which has   motivated their widespread use in the automotive, marine, and aerospace industries, primarily as lightweight cores in sandwich  panels \cite{corigliano2000experimental,li2008self}.   Syntactic foams have also found widespread use in other  applications, particularly in thermal management and  for vibration isolation \cite{gupta2014applications,hu2011tensile,banea2014mechanical}. In mechanical applications, the vast majority of syntactic foams comprise  hollow glass microspheres (HGM) as a filler \cite{
d1999analysis,
gupta2001studies,
kim2004manufacturing,
gupta2006characterization,
gupta2010comparison,
salleh2013preliminary,
lachambre2013situ,
pellegrino2015mechanical,
pinisetty2015hollow,
zeltmann2017mechanical}, as they increase the small-strain stiffness and compressive yield strengths of the matrix  \cite{pinisetty2015hollow}. However, HGM syntactic foams are often heavier than HTM syntactic foams at the same filling fraction and for certain matrix materials, such as glassy matrices, possess poor mechanical recoverability (i.e., the stiffness does not recover on subsequent load cycles). 

An emerging class of syntactic foam   comprises hollow thermoplastic microspheres (HTMs) embedded in a polymer matrix,  which have   demonstrated   increased small-strain stiffness,   lower densities, and    strong recovery    at large deformations     \cite{everett1997preliminary,everett1998microstructure,kim2009toughening,banea2014mechanical,banea2015structural,dando2020nano}.   The property of  strong (stiffness) recoverability is a feature  attributed to the elastic buckling response of individual  copolymer shells \cite{shorter2010axial}, which contrasts syntactic foams made with glass microspheres that typically crack under large strains. The recoverability is also in part assisted by the matrix material choice, with soft elastomers like polyurethane having an advantage over those that are   more brittle, such as epoxy resins; recently the introduction of HTMs has been found to improve the recoverability of   epoxy at high filling fractions  \cite{dando2019characterization,dando2018production}  .

In this work,        the   mechanical   performance of elastomeric   HTM syntactic foams is investigated. Earlier works on the mechanical properties of polyurethane-HTM syntactic foams   \cite{everett1997preliminary,everett1998microstructure} have  focused    on  low-filling fraction samples,  subsequently missing key features of these foams. The present study steps outside this regime and systematically examines   transitions across a range of filling fractions  from $\phi=0\%$ to $\phi = 40$\%, complementing and extending preliminary work   by some of the authors \cite{yousaf2020compression}. By conducting a thorough experimental characterisation of the mechanical and thermal properties, and by presenting phenomenological models for the elastic response (contributing to the limited modelling literature that is presently available for HTM syntactic foams \cite{de2013predicting,shrimali2020simple,paget2020syntactic}),   it is anticipated that the   present study will greatly assist ongoing materials and device development. For example, the strain energy models provided may be of use in  numerical predictions describing the mechanical performance of layered media or  sandwich structures with a syntactic foam core, as well as in the digital design of novel materials  via machine learning \cite{gu2018novo,chen2019machine,pal2021machine}.

 Our work reveals an   {increased}  small-strain stiffness and {reduced} density with increasing filling fraction. The syntactic foams are  also found to  possess strong stiffness recoverability (i.e., minimal permanent softening) to large peak strains. Strain energy   models are   constructed to describe the stress-strain and transverse mechanical response        \cite{ogden1997non}, which  are   qualitatively well-described by   simplified Ogden models \cite{ogden1997non,ogden1972large} at low filling fractions and    by advanced Ogden models at large filling fraction \cite{ogden1982elastic,schrodt2005hyperelastic}. When modelling HTM syntactic foams, a significant challenge lies in finding      strain energy ansatzes that capture   both the   axial stress-strain and transverse strain response of the syntactic foam, requiring an appropriate compressibility condition $f(J)$ in the strain energy \cite{ogden1997non,schrodt2005hyperelastic}. We find that outside the dilute regime (i.e., $\phi>10\%$) the compressibility condition must be suitably general for HTM syntactic foams. 
 
 We close this work by briefly exploring the thermal characteristics of these foams, and  find that the presence of HTMs does not significantly modify the thermal properties of the polyurethane matrix. These  thermal results  contrast   those for HGM syntactic foams in epoxy  \cite{lin2009thermoanalytical}, but is similar to HGM syntactic foams in polystyrene and PMMA where the impact is also minimal \cite{li2008self,ozkutlu2018effects}. This finding motivates the use of HTMs as a lightweight substitute (filler)  for polyurethane in thermal applications, which when coupled with their enhanced mechanical performance, demonstrates the highly multipurpose nature of HTM syntactic foams.

\section{Material Fabrication and Mechanical Testing} 
\label{sec:expsetup}

\begin{figure}[t]

\centering
\subfigure[  \label{fig:92025p} ]{\includegraphics[width=0.304\textwidth]{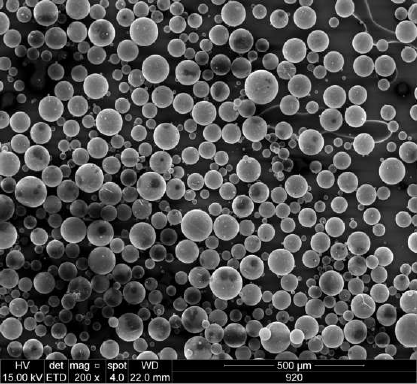}}
\subfigure[  \label{fig:92025b} ]{\includegraphics[width=0.304\textwidth]{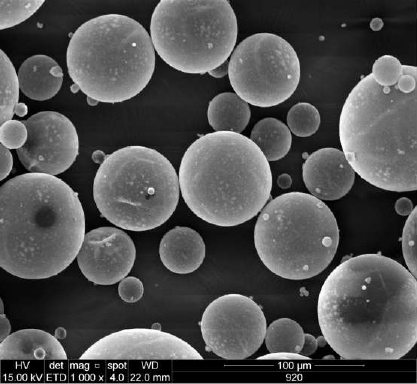}} \\
\subfigure[  \label{fig:sem92040a} ]{\includegraphics[width=0.304\textwidth,angle=0]{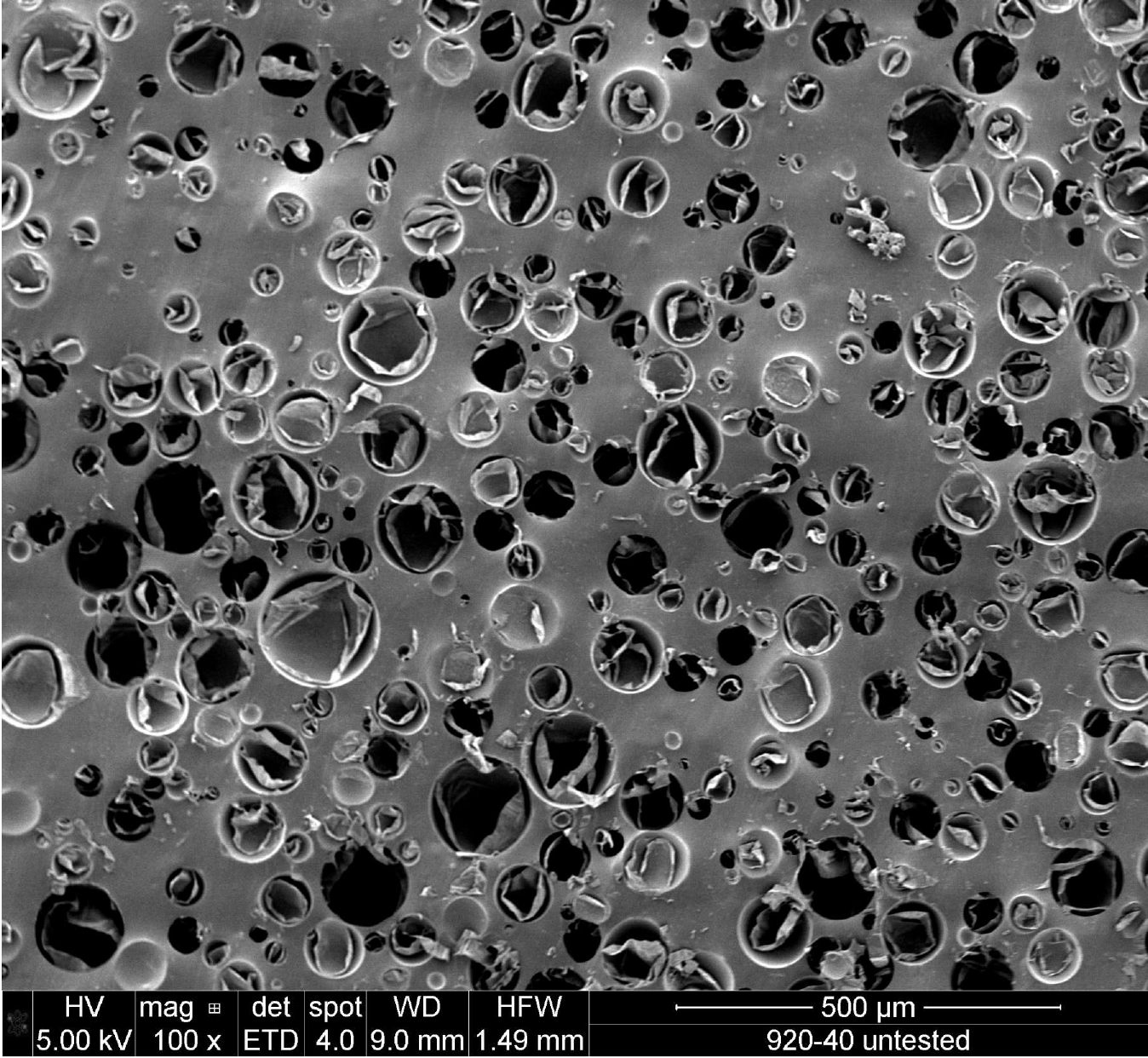}}
\subfigure[  \label{fig:sem92040b} ]{\includegraphics[width=0.304\textwidth,angle=0]{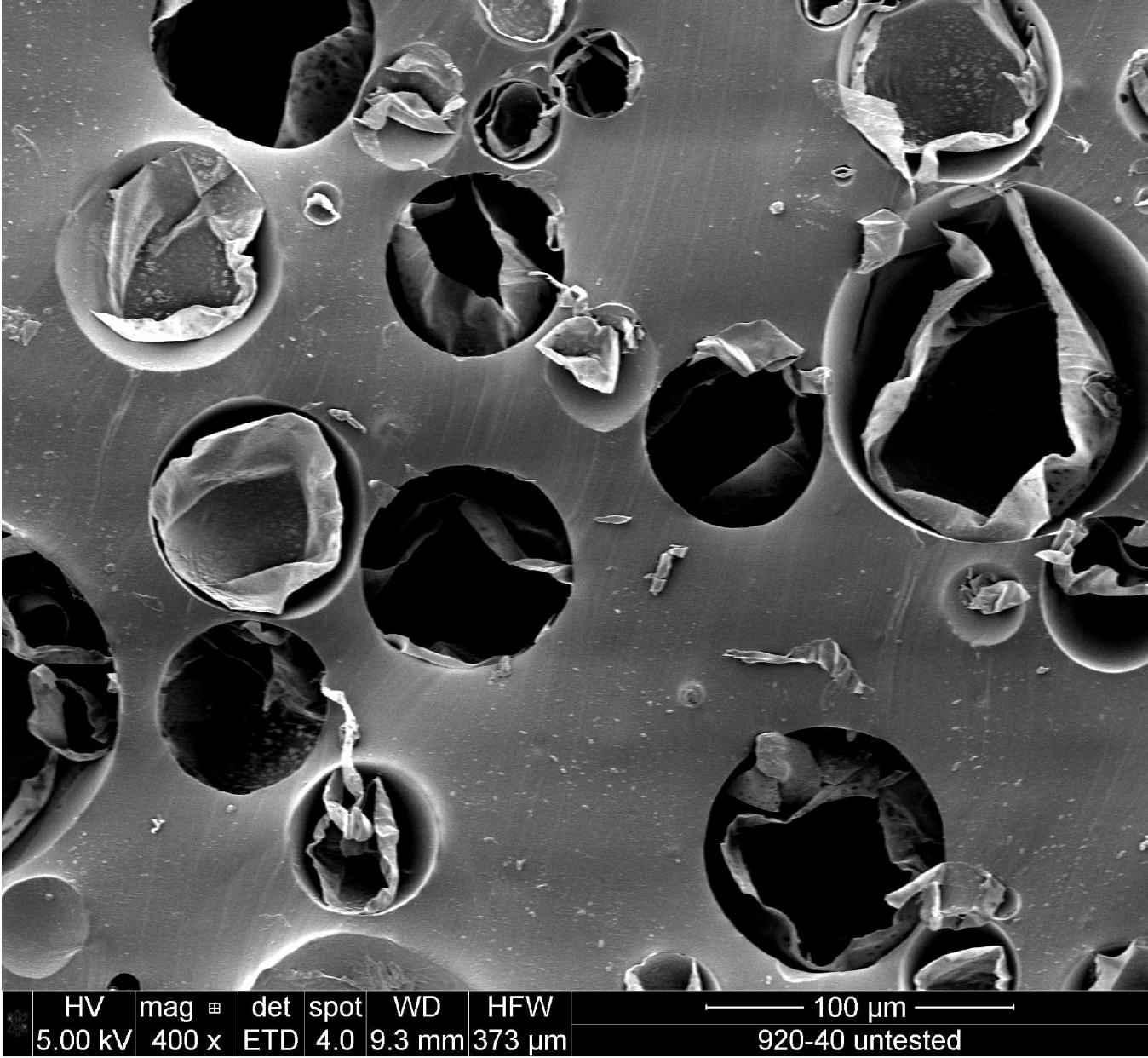}}
\caption[Scanning Electron Microscope (SEM) images at different magnifications of \protect\subref{fig:92025p},\protect\subref{fig:92025b} unsuspended hollow thermoplastic microspheres;  \protect\subref{fig:sem92040a},\protect\subref{fig:sem92040b}  HTM foam sample  surface    at $\phi=40\%$ filling fraction.]{(Caption in list of figures) \label{fig:SEM}} 

\end{figure}
\subsection{Materials and sample preparation}
The HTM syntactic foams   were made by blending hollow copolymer microspheres (Expancel 920 DE supplied by Expancel AzkoNobel)   into a polyurethane   matrix   made from a blend of
 Polytetramethylene Ether Glycol  
(Terathane 1000 supplied by INVISTA Textile (UK) Ltd), Trimethylolpropane  (supplied by Tokyo Chemical
Industry) with   Methylene diphenyl diisocyanate (Isonate M143 supplied by Dow Chemicals) as a curing agent.
Fumed silica (Aerosil 200 supplied by Evonik Inc.) was used as a thixotropic additive.   A summary of  the microsphere properties  is given in Table \ref{tab:table1}, where we note that   particle diameter values  are volume-weighted   \cite{curd2020characterisation}.  Scanning-electron microscope (SEM) images of the spheres  are given in Fig.~\ref{fig:SEM}, both in suspension and    resting on the imaging surface.  After blending, the mixture was cured   in open   trays at 55$^\circ$C and then machined into    cylinders (diameters of 29 mm and heights of 12.5 mm).  

\begin{figure}[t]

\centering
\includegraphics[width=0.6\textwidth]{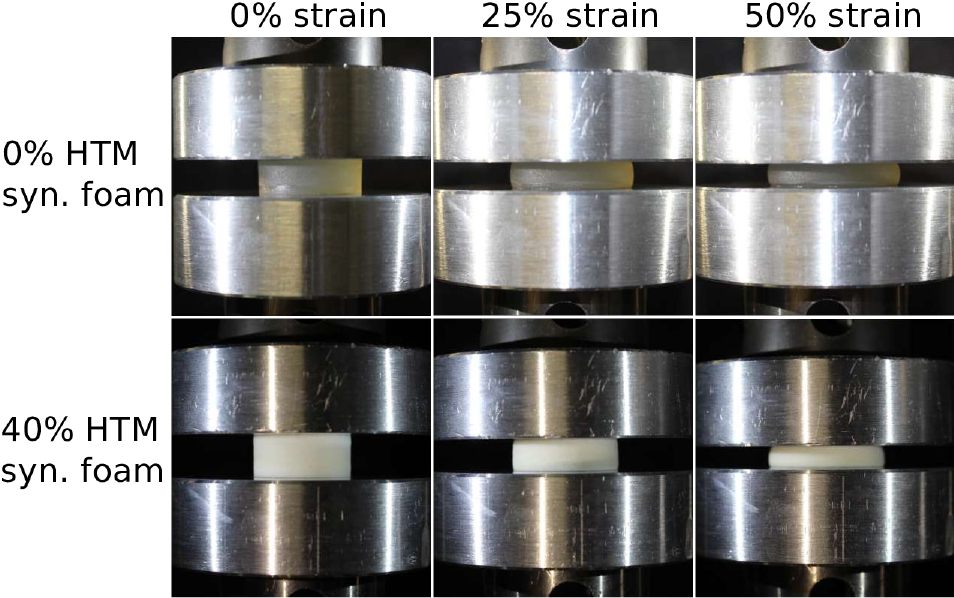}
 
 \caption[Test setup for measuring compressive   performance.]{(Caption in list of figures) \label{fig:fig3}}
 
\end{figure}

\subsection{Mechanical testing methods}
Transversally unconfined uniaxial compression testing  of the samples was conducted on an Instron     testing machine   (see Fig.~\ref{fig:fig3}) at  a strain rate of  10 mm/min, following    BS ISO 7743-2011.   The top and bottom platens were sprayed with WD-40  to minimise   barrelling of the samples due to friction between the platens and the sample.   The relationship between the lateral (radial)   and axial stretch     under loading was determined from video recordings   of the samples under compression. Further details are given in the Supplementary Material on extracting the Poisson response of the samples.

\section{Experimental results and discussion} 
\label{sec:compr}
In all figures,     results are given for  the   initial loading curve    at room temperature averaged over  2-3 samples. For reference, the raw and toe-compensated    experimental data  generated in connection with this work is  freely available on   FigShare   \cite{smith2018analyticaldata}.

\begin{figure}[t]

\centering

 \subfigure[  \label{fig:ss70920} ]{
\includegraphics[width=0.37\textwidth]{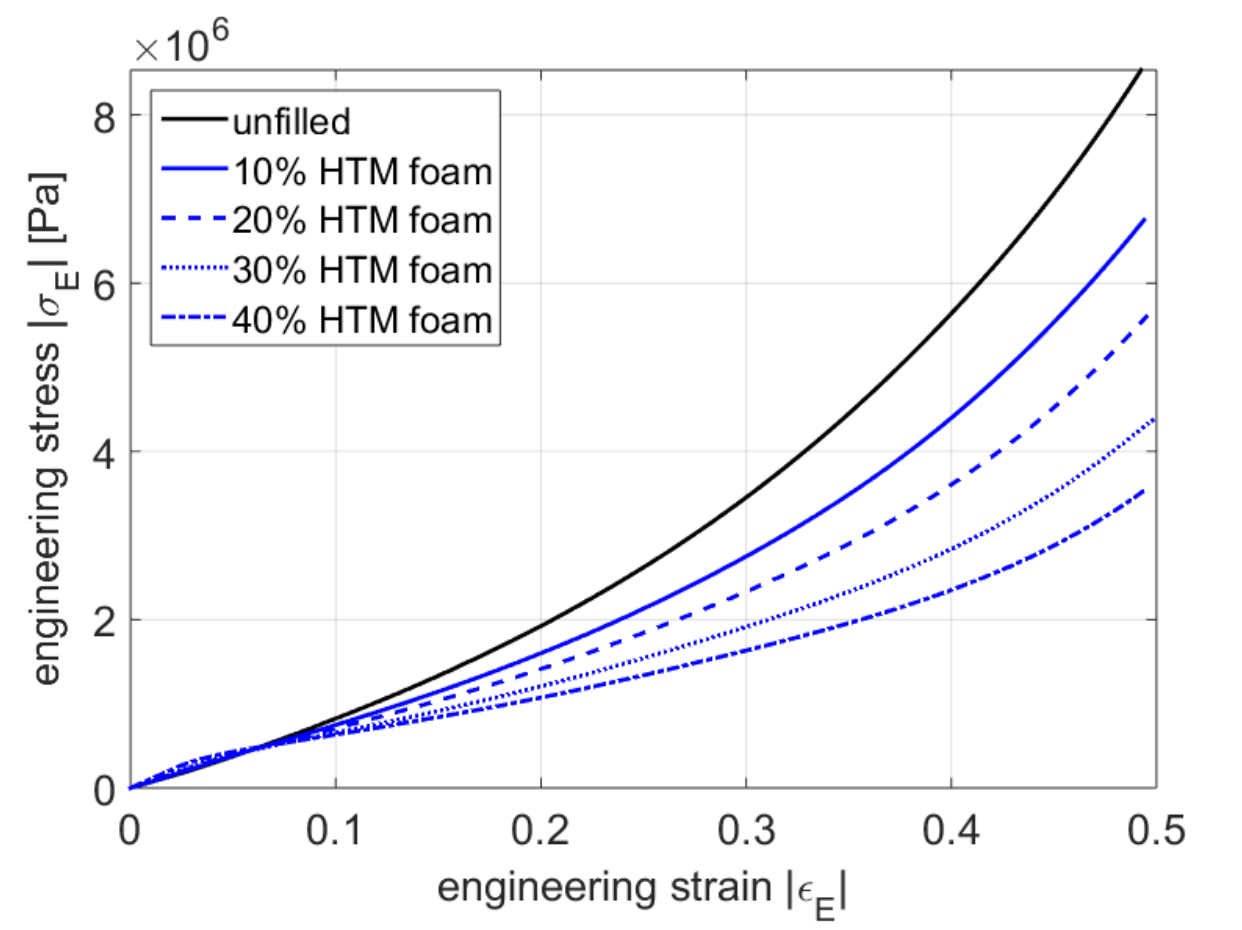}
 }
   \subfigure[  \label{fig:ss25920} ]{
\includegraphics[width=0.37\textwidth]{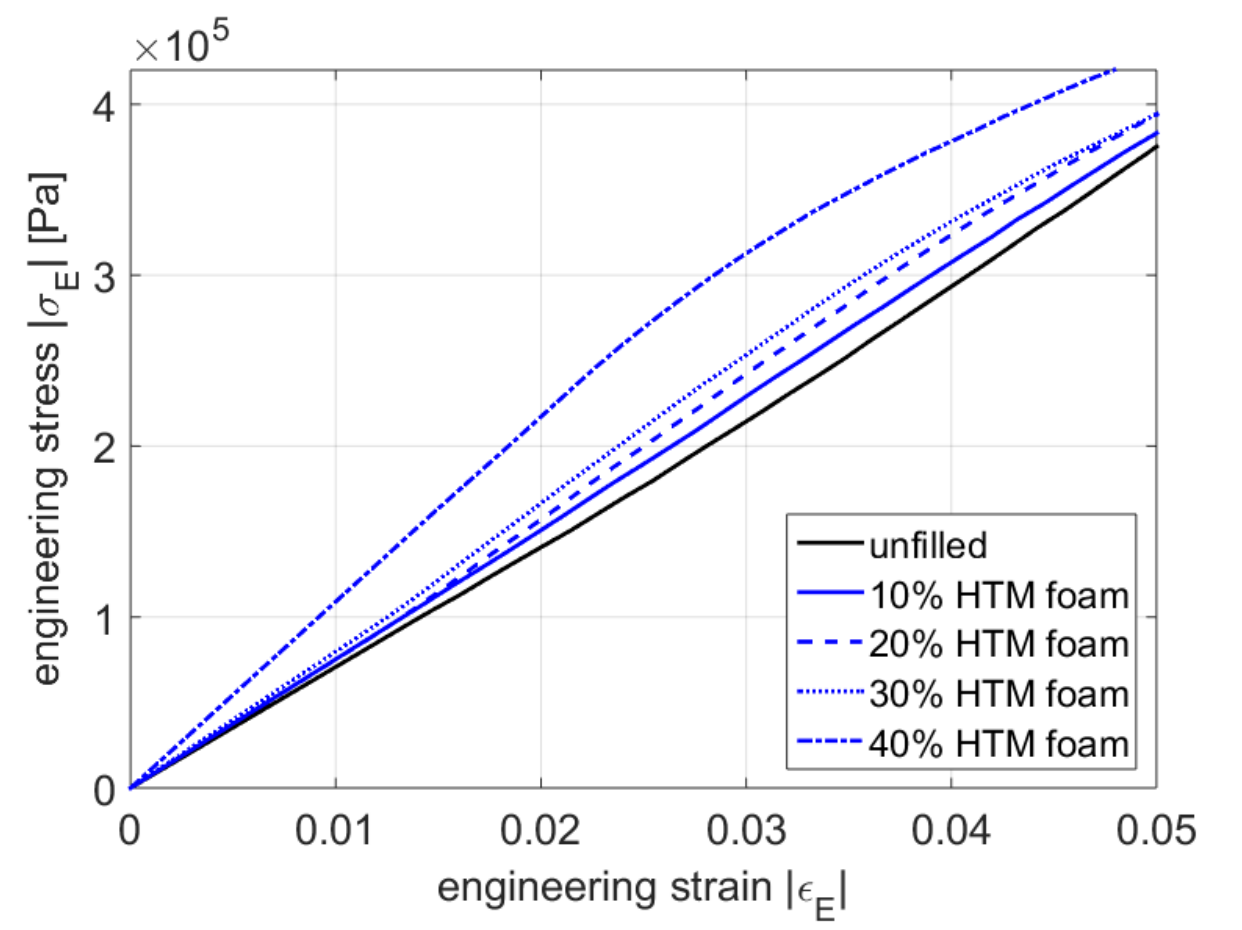}
 }
\caption[Stress-strain test data    up to \subref{fig:ss70920} $50\%$   strain; \subref{fig:ss25920}   5\% strain (zoom-in).]{(Caption in list of figures)   \label{fig:70percent920} }

  \end{figure}
\subsection{Axial stress-strain response}
 Fig.~\ref{fig:70percent920} shows the axial stress-strain diagram for HTM syntactic foams  at the  filling fractions $\phi = 0\%$ (unfilled),  10\%, 20\%, 30\%, and 40\%,  up to compressive strains of 50\%.  We observe that   the   samples    exhibit  nonlinear elastic behaviour typical of a polymeric material \cite{ogden1997non} at dilute  filling fractions $\phi \leq 10\%$. At non-dilute filling fractions,   the samples   exhibit       conventional syntactic foam behaviour \cite{gibson1999cellular}: the emergence of a linear region at small strains, a graded plateau region at medium strains, and strong densification at large strains.   Fig.~\ref{fig:ss25920} presents a zoom-in of Fig.~\ref{fig:ss70920} up to 5\% strain, clearly demonstrating   small-strain stiffening   effects. The results for $\phi >10\%$     suggest  that  interaction effects between closer-packed HTMs, in addition to the buckling response of individual HTMs,    significantly influence the macroscale  foam response.   

\begin{figure}[t]

\centering

 \subfigure[  \label{fig:lzlr920} ]{
\includegraphics[width=0.37\textwidth]{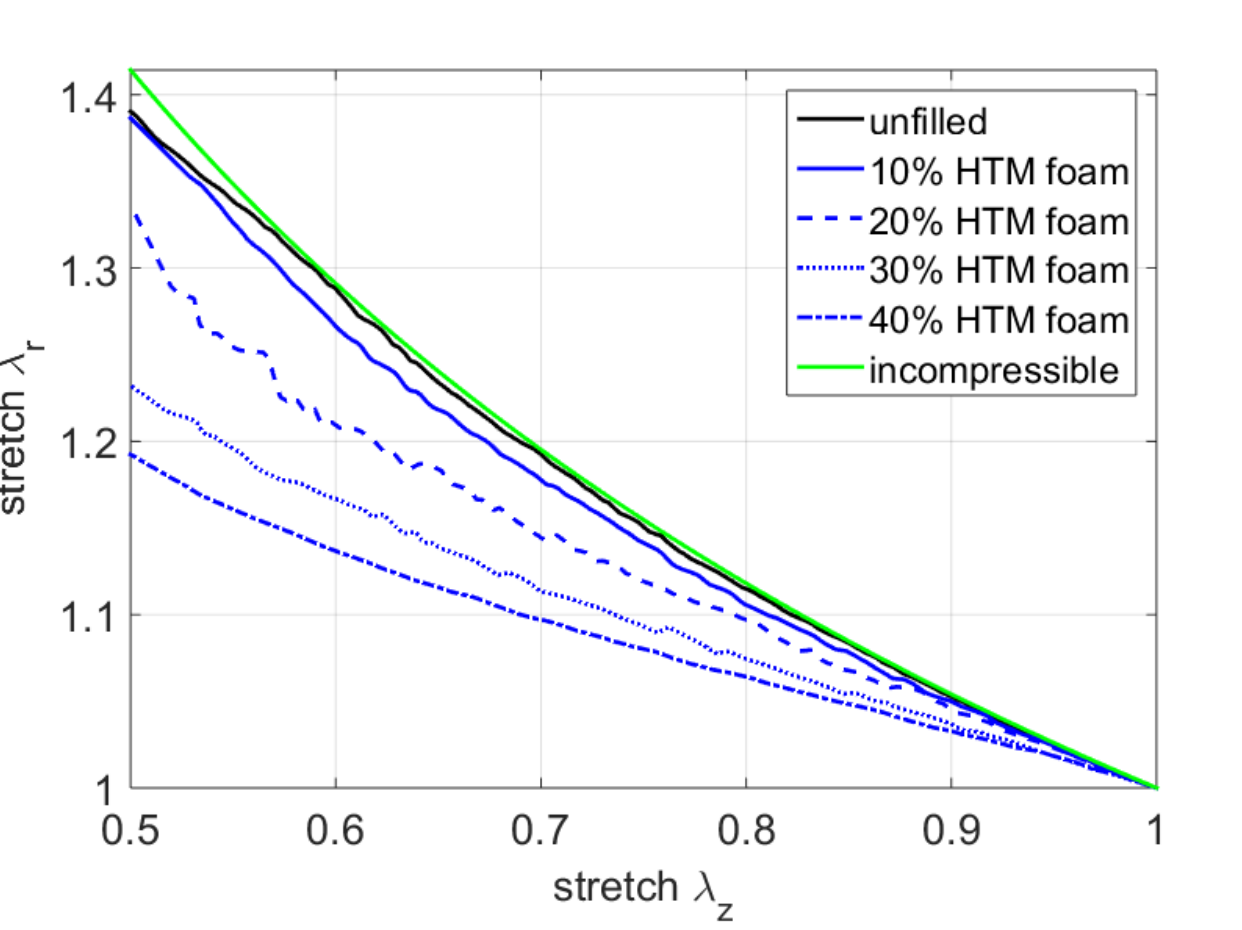}
}
 \subfigure[  \label{fig:lzJ920} ]{
\includegraphics[width=0.37\textwidth]{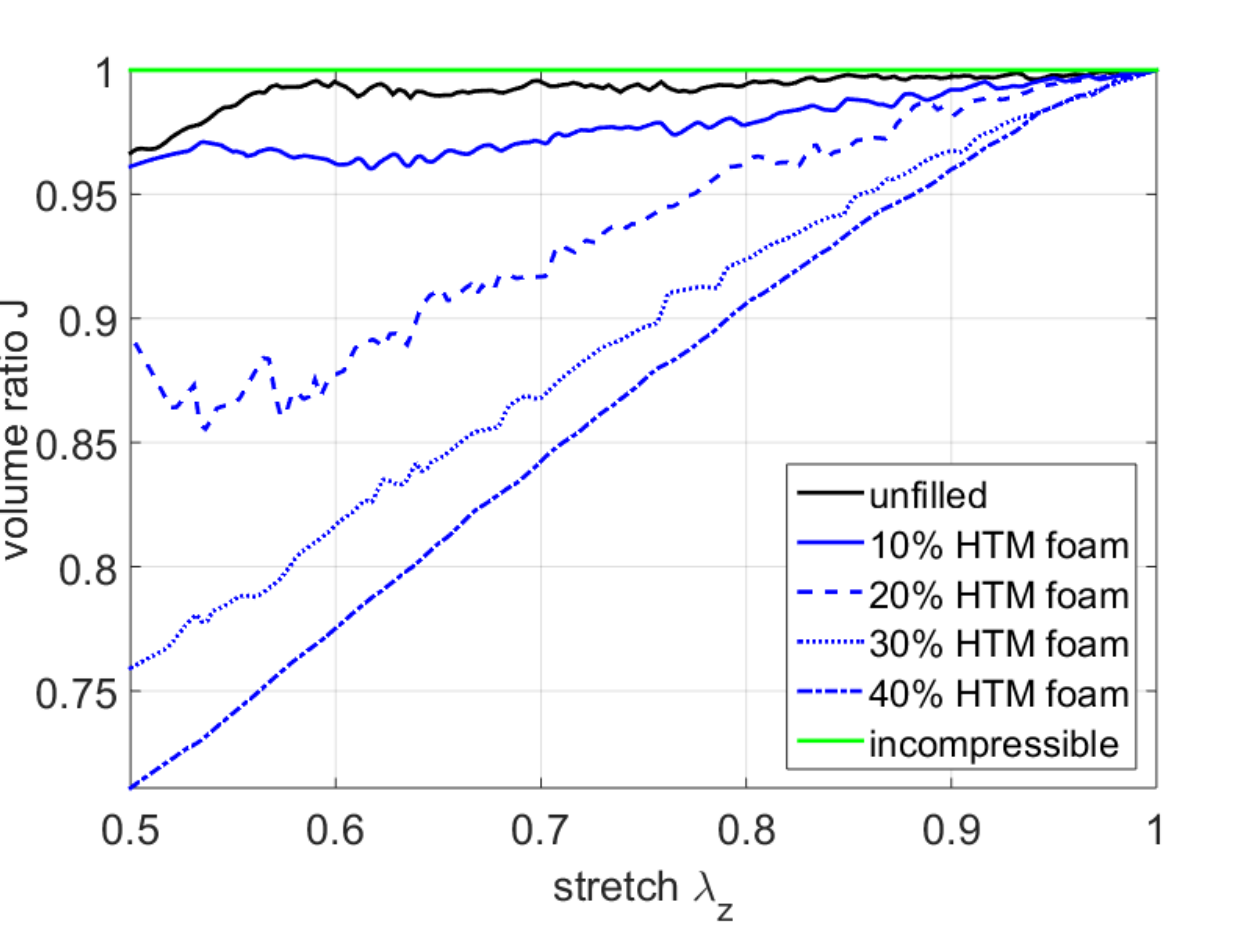}
}
\caption[Transverse stretch test data     up to $50\%$   strain: \protect\subref{fig:lzlr920}      radial stretch $\lambda_\rmr$  as   function of axial stretch $\lambda_\rmz$;  \subref{fig:lzJ920} volume ratio       $J = V/V_0 = \lambda_\rmr^2 \lambda_\rmz$.]{(Caption in list of figures) \label{fig:transversedatafigs}   }
 
  \end{figure}
\subsection{Transverse-to-axial strain response}
The corresponding transverse-to-axial strain response of these materials under   compression is   given in Fig.~\ref{fig:lzlr920}. Results are expressed in terms of   stretches    $\lambda$, defined via the engineering strain $\epsilon_\mathrm{E} = \lambda - 1$, where $\lambda_\rmr$ and $\lambda_\rmz$ denote the  radial   and axial stretch, respectively.   Fig.~\ref{fig:lzlr920}   presents test data for $\lambda_\rmr$ as a function of $\lambda_\rmz$ where     dilute $\phi$  foams  respond   comparably to an incompressible material  (green reference curve). For non-dilute foams,   a considerable change in the   response was observed; we observe the Poisson response becoming a function of strain, which for the case of $\phi = 40\%$ shows a $15\%$ reduction in $\lambda_\rmr$ at $50\%$ strain relative to the unfilled material. The highly compressible response of the foam is attributed to both  the strong compressibility of the enclosed   gas within the HTMs  and the buckling response of the microsphere shells.

\subsection{Volume ratio response}
  Fig.~\ref{fig:lzJ920} gives the corresponding volume ratio $J = \lambda_\rmr^2 \lambda_\rmz$, where   low filling fraction samples correspond to     weakly   compressible  media ($J\approx1$).  However for $\phi=40\%$,          significant compressions    of  $J \approx 0.75$   at 50\% strain were observed, in addition to the emergence of   linear and large-strain response regions as discussed above in Fig.~\ref{fig:70percent920}. This volume ratio response demonstrates a considerable increase in the compressibility relative to the matrix material.

\begin{figure}[t]

\centering
\subfigure[  \label{fig:youngsmod} ]{
\includegraphics[width=0.33\textwidth]{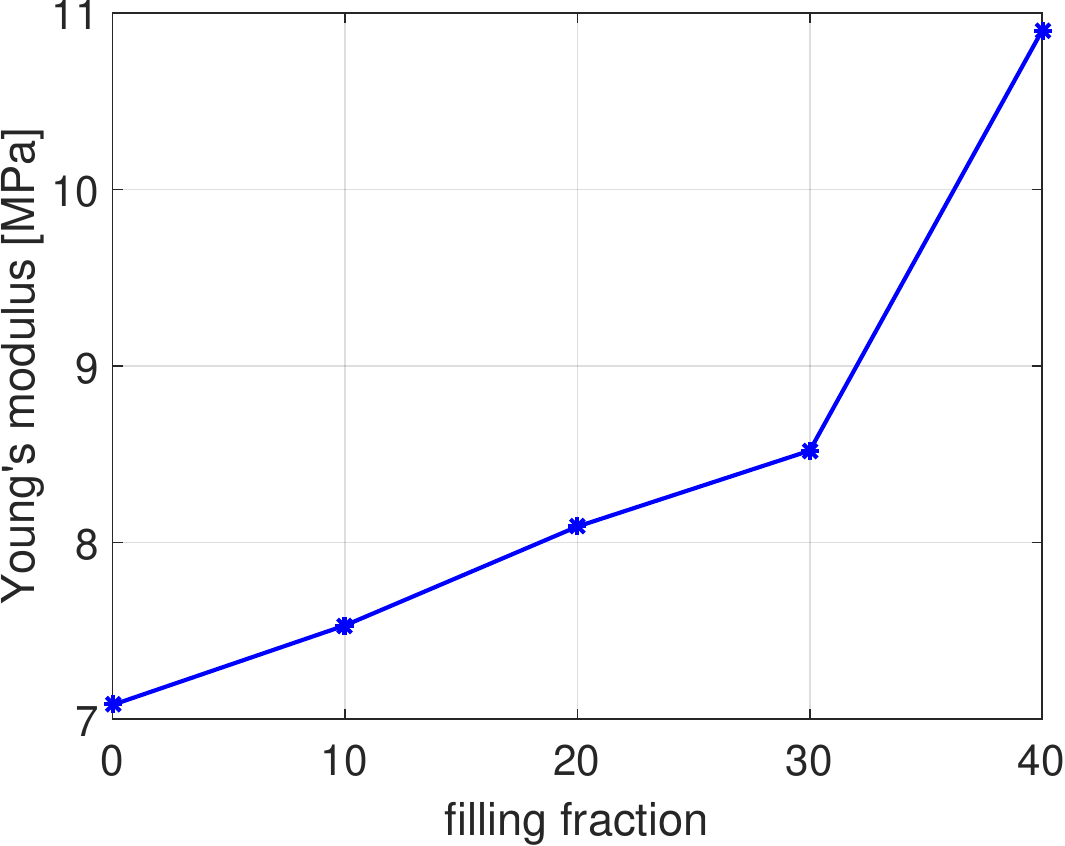}
}
 \subfigure[  \label{fig:poissonrat} ]{
\includegraphics[width=0.33\textwidth]{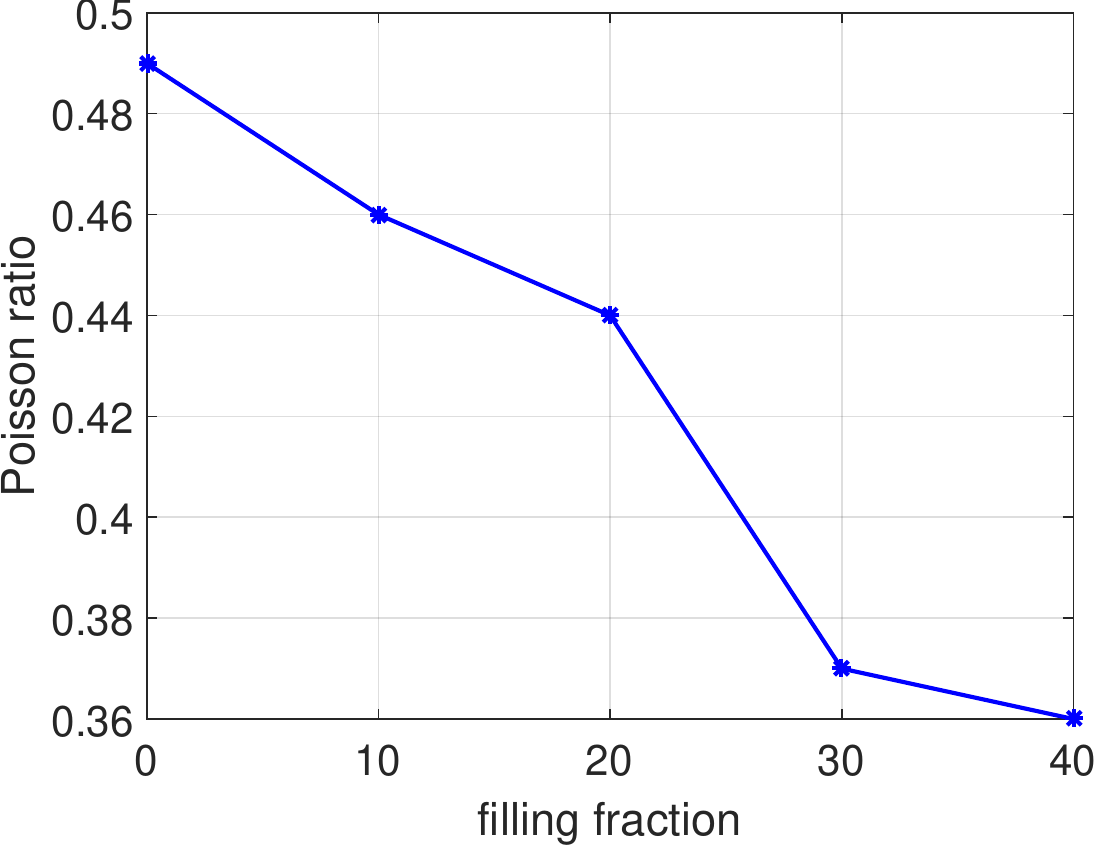}
}

 \caption[  Experimental   values for \subref{fig:youngsmod} Young's modulus and \subref{fig:poissonrat} Poisson's ratio of syntactic foam samples against filling fraction]{(Caption in list of figures) \label{fig:youngpoisson}}
 
\end{figure}

 \subsection{Small-strain materials constants}
 Table \ref{tab:smstrconsts} presents the measured small-strain properties of the   syntactic foams, namely the  Young's modulus $Y$, and the   Poisson ratio $\nu$ (both calculated   at approximately $2\%$ strain),  following the procedure discussed   in Sec.~\ref{sec:expsetup}.  The measured density $\rho$ and corresponding specific stiffness $Y/\rho$ of these materials is also included,   revealing an enhancement in the specific stiffness of the   matrix, for example, by a factor  of   $2.5$ at $\phi = 40\%$, and thus reveals an extremely lightweight   alternative to other syntactic foams   \cite{d1999analysis}. A further discussion of the density is given below. Results for the Young's modulus and Poisson's ratio are presented in Fig.~\ref{fig:youngpoisson} where a significant enhancement in the small-strain stiffness is observed for $\phi>30\%$ and the transition from polymeric to  syntactic foam behaviour is indicated for $\phi>20\%$ from the Poisson ratio data.

\begin{figure}[t]

\centering
\subfigure[  \label{fig:density} ]{
\includegraphics[width=0.33\textwidth]{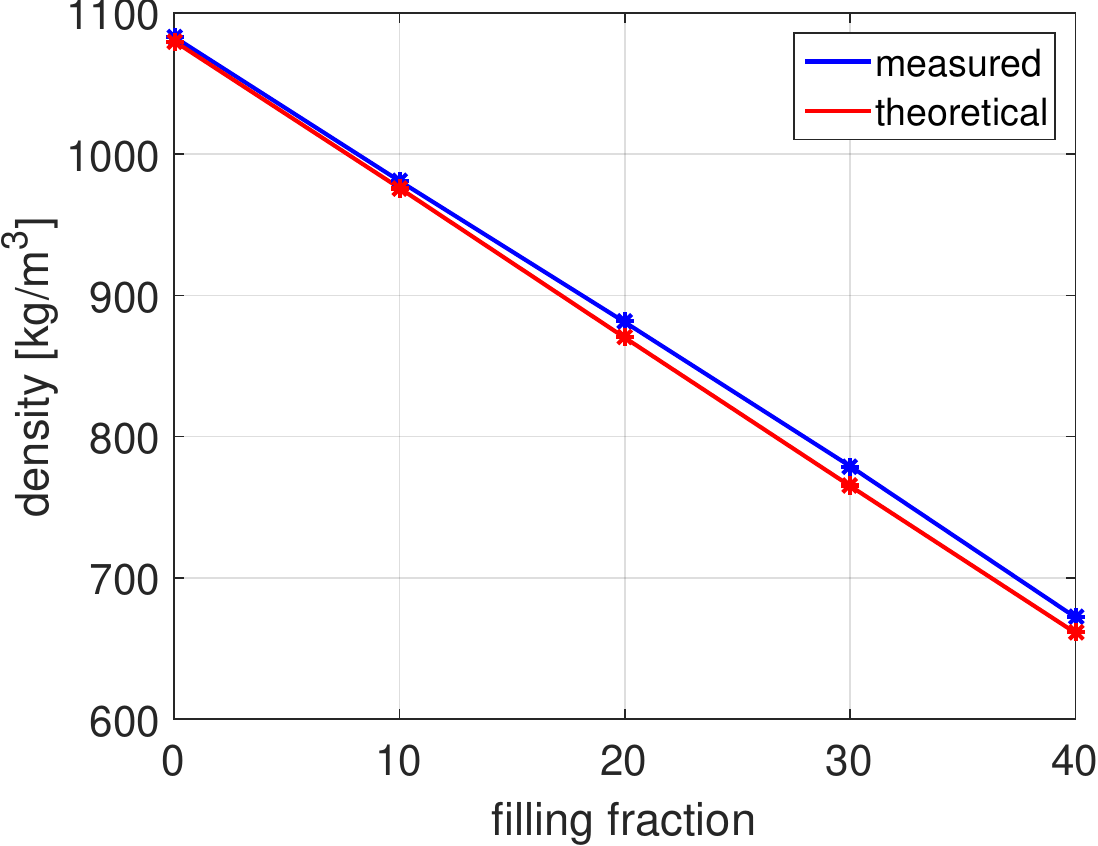}
}
 \subfigure[  \label{fig:specificstiff} ]{
\includegraphics[width=0.33\textwidth]{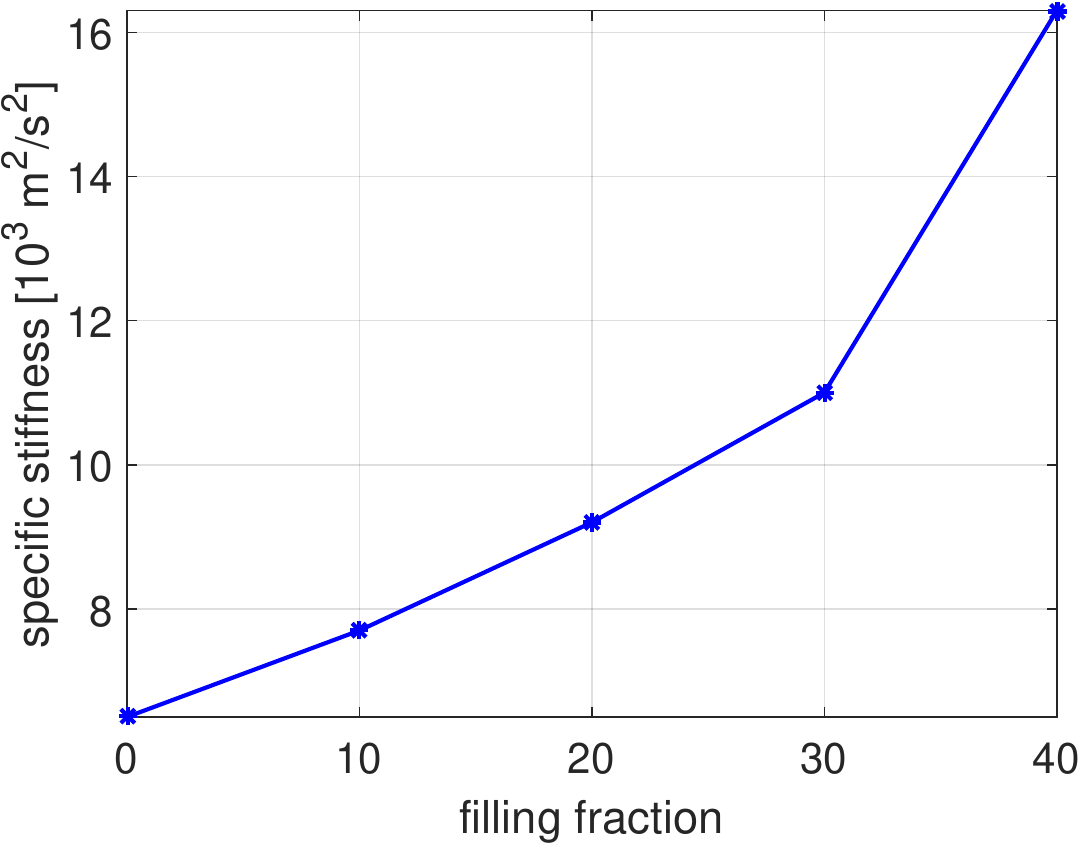}
}

 \caption[ \subref{fig:density} Experimental and theoretical values for density of syntactic foam samples against filling fraction, \subref{fig:specificstiff} Experimental values for corresponding specific stiffness]{(Caption in list of figures) \label{fig:figdensityandspecstiff}}
 
\end{figure}
 \subsection{Density} 
Fig.~\ref{fig:density}  presents the measured and theoretical density values for all syntactic foam samples, in accordance with the results presented in Table~\ref{tab:smstrconsts}. Here we observe that the theoretical curve given by $\rho_\mathrm{th} = \rho_\mathrm{mat}(1-\phi) + \rho_\mathrm{mb}  \phi$, where $\rho_\mathrm{mat} = 1083$ kg/m$^3$ is the density of the matrix and $\rho_\mathrm{mb} = 28.7$ kg/m$^3$ is the measured density of the microballoons \cite{daniel2017correspondence}, closely matches the curve formed from measured values. This suggests that manufacturing defects are well-controlled across all samples. For reference,  Fig.~\ref{fig:specificstiff} gives the corresponding specific stiffness for all filling fractions showing a significant enhancement, as discussed above.

   \begin{figure}[t]
 
\centering

\subfigure[  \label{fig:poisson1} ]{
\includegraphics[width=0.375\textwidth]{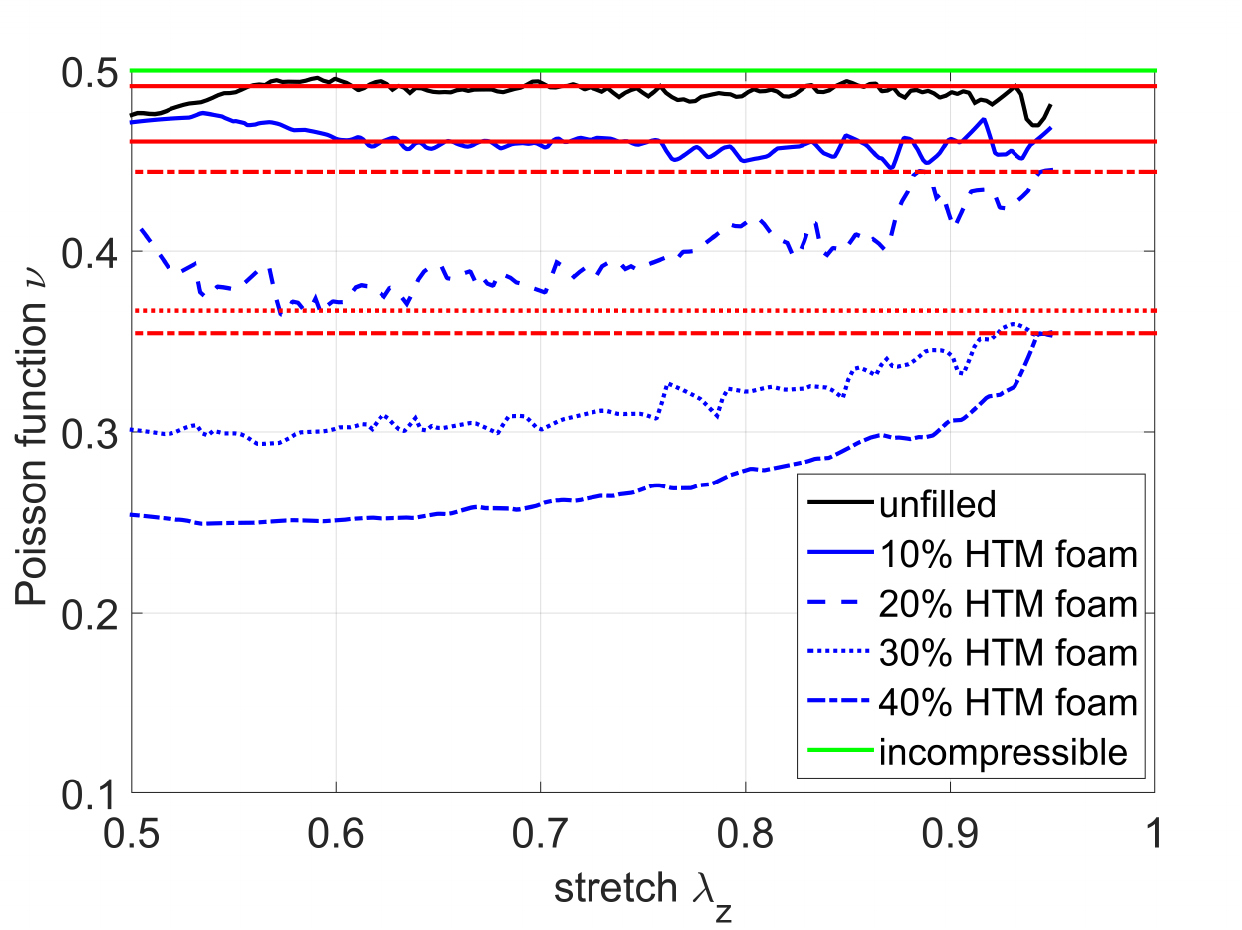}
}
 \subfigure[  \label{fig:poisson2a} ]{
\includegraphics[width=0.33\textwidth]{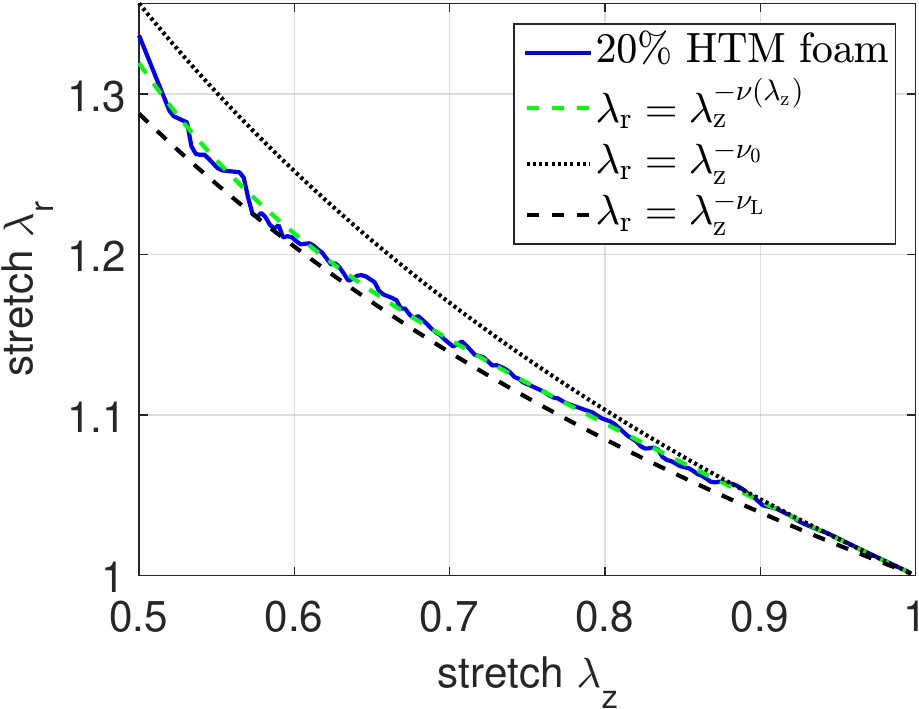}
}
\\
 \subfigure[  \label{fig:poisson2b} ]{
\includegraphics[width=0.33\textwidth]{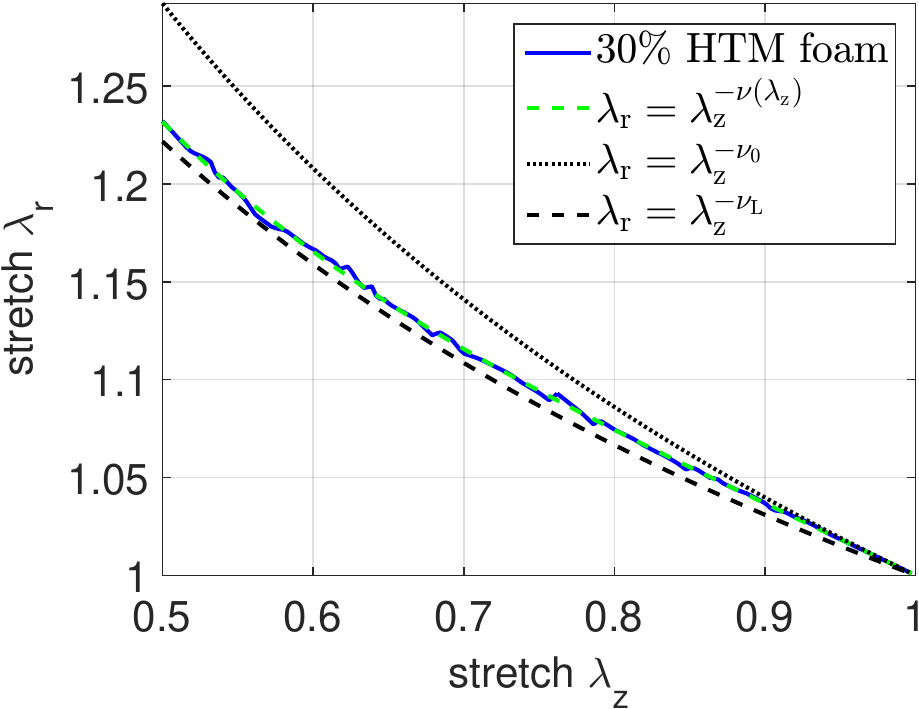}
}
 \subfigure[  \label{fig:poisson2} ]{
\includegraphics[width=0.33\textwidth]{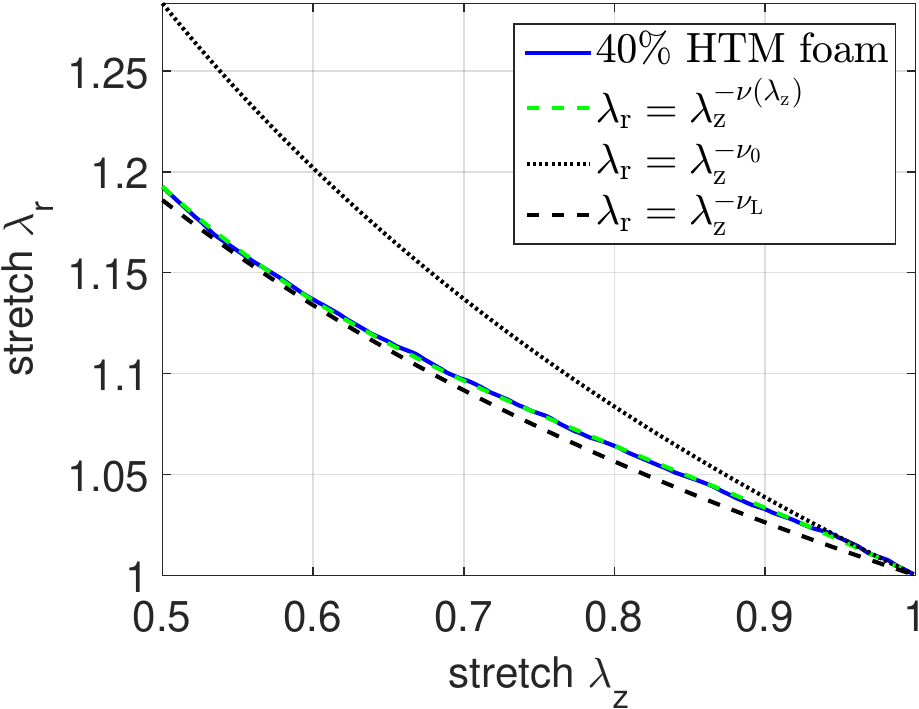}
}

\caption[Test data       up to 50\%   strain:  \subref{fig:poisson1} Poisson ratio $\nu$ as a function of $\lambda_\rmz$   with $\nu_0$   overlaid  (red curves); \subref{fig:poisson2}     simplified models for transverse stretch behaviour   of $\phi =  20\%, \, 30\%, \, \mbox{and}\, 40\%$  foams.]{(Caption in list of figures) \label{fig:poissonsl92055140}}

  \end{figure}
 \subsection{Poisson function}
Fig.~\ref{fig:poisson1}  presents the   Poisson function for all filling fractions,  defined in terms of   the   {\it Hencky ratio}   \cite{mihai2017characterize}  
\begin{equation}
\label{eq:linmodellrlz}
\nu(\lambda_\rmz) \approx - {\log(\lambda_\rmr)}/{\log(\lambda_\rmz)}.
\end{equation}
Also superposed is  the Hencky form     $\lambda_\rmr \approx \lambda_\rmz^{-\nu_0}$ for the transverse response, where  $\nu_0$ is the small-strain Poisson's ratio presented in  Table \ref{tab:smstrconsts}.  We observe a reduction in both the Poisson ratio and Poisson function with increased filling fraction (i.e., increasing compressibility with increased filling fraction), as well as a clear dependence on stretch at higher filling fractions. A Poisson  ratio accurately describes the   transverse response   for very dilute $\phi$,  as shown by the straight line fits   (red curves). However  for $\phi > 10\%$, the transverse response  is a function of   strain; we find that the response is qualitatively well-described  by    
 \begin{equation}
\label{eq:modellzlr}
\lambda_\rmr \approx \lambda_\rmz^{-\nu(\lambda_\rmz)} , \quad \mbox{where} \quad  \nu(\lambda_\rmz) \approx \kappa_3 (\lambda_\rmz - 1)^3 + \kappa_2 (\lambda_\rmz - 1)^2   +\kappa_1 (\lambda_\rmz - 1)  + \kappa_0, 
\end{equation} 
with   $\kappa_0 \approx \nu_0$.    Figs.~\ref{fig:poisson2a}--\ref{fig:poisson2}    examines   these models for $\phi =20\%$, $30\%$ and $40\%$ filling fraction foams, superposing the   test data (blue line)   with   Eq.~\eqref{eq:modellzlr}  where  $(\kappa_3,\kappa_2,\kappa_1,\kappa_0) =( -2.1454,   -1.1894,   0.0218,    0.44)$, 
 $(\kappa_3,\kappa_2,\kappa_1,\kappa_0) =(-0.1252,    0.2591,    0.2991,    0.37)$, and
 $(\kappa_3,\kappa_2,\kappa_1,\kappa_0) =(0.0710,    0.6966,    0.5421,    0.36)$ (green dashed lines), respectively.  Reference curves for  the Hencky ratio \eqref{eq:linmodellrlz} at   small- and large- strains    
 ($\nu_0  \approx 0.44$ and $\nu_\mathrm{L}  \approx 0.37$, 
$\nu_0  \approx 0.37$ and $\nu_\mathrm{L}  \approx 0.3$, and
 $\nu_0  \approx 0.36$ and $\nu_\mathrm{L}  \approx 0.25$, 
 respectively) are also superposed. We remark that our   Poisson function form   \eqref{eq:linmodellrlz} and transverse response form \eqref{eq:modellzlr} are by no means unique, with other authors combining Biot and Hencky measures to successfully describe the Poisson response, for example (see \cite{beatty1986poisson,ciambella2014continuum,mihai2017characterize}).

\begin{figure}[t]

\centering

\subfigure[  \label{fig:92025prec} ]{
\includegraphics[width=0.375\textwidth]{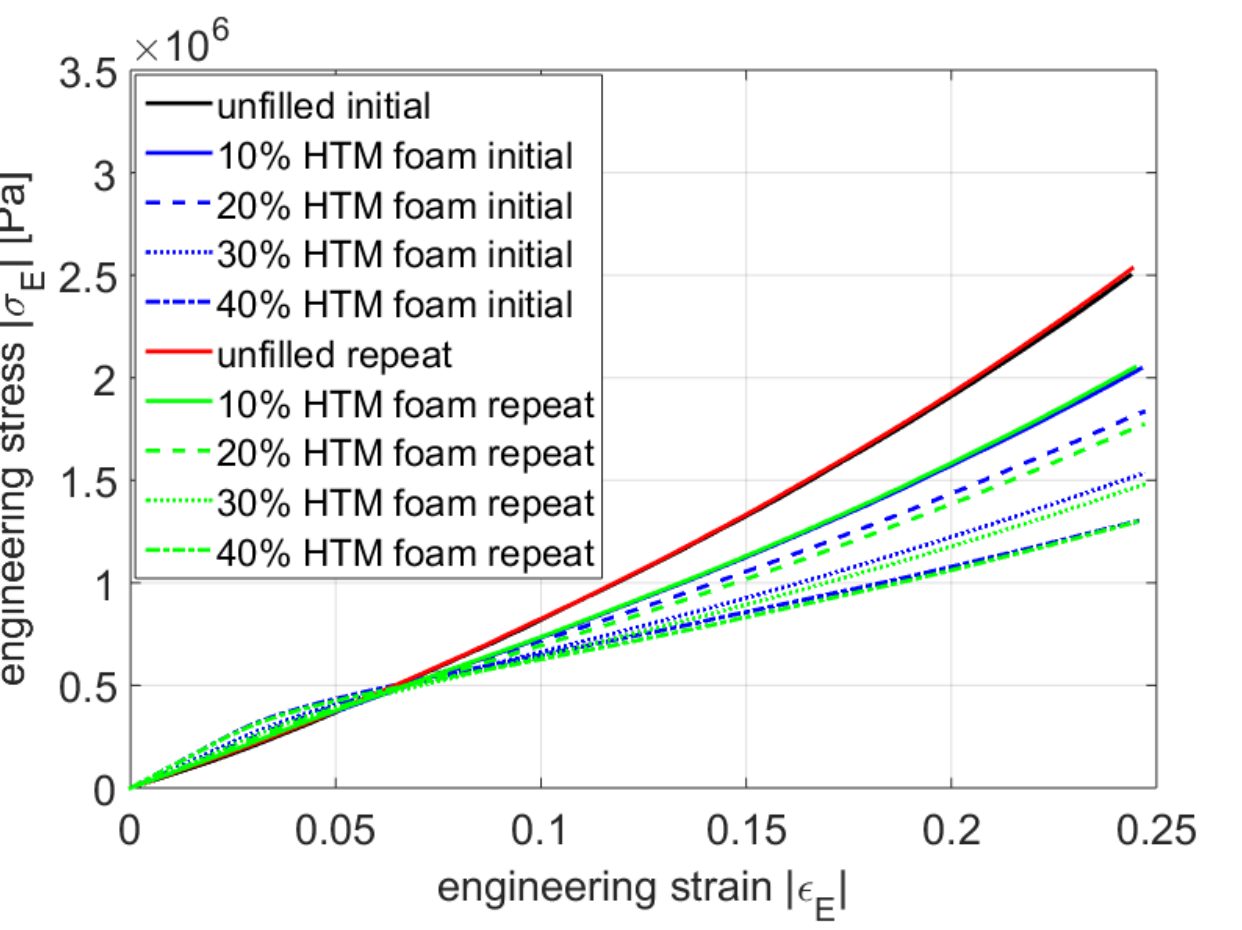}
}
 \subfigure[  \label{fig:92050prec} ]{
\includegraphics[width=0.375\textwidth]{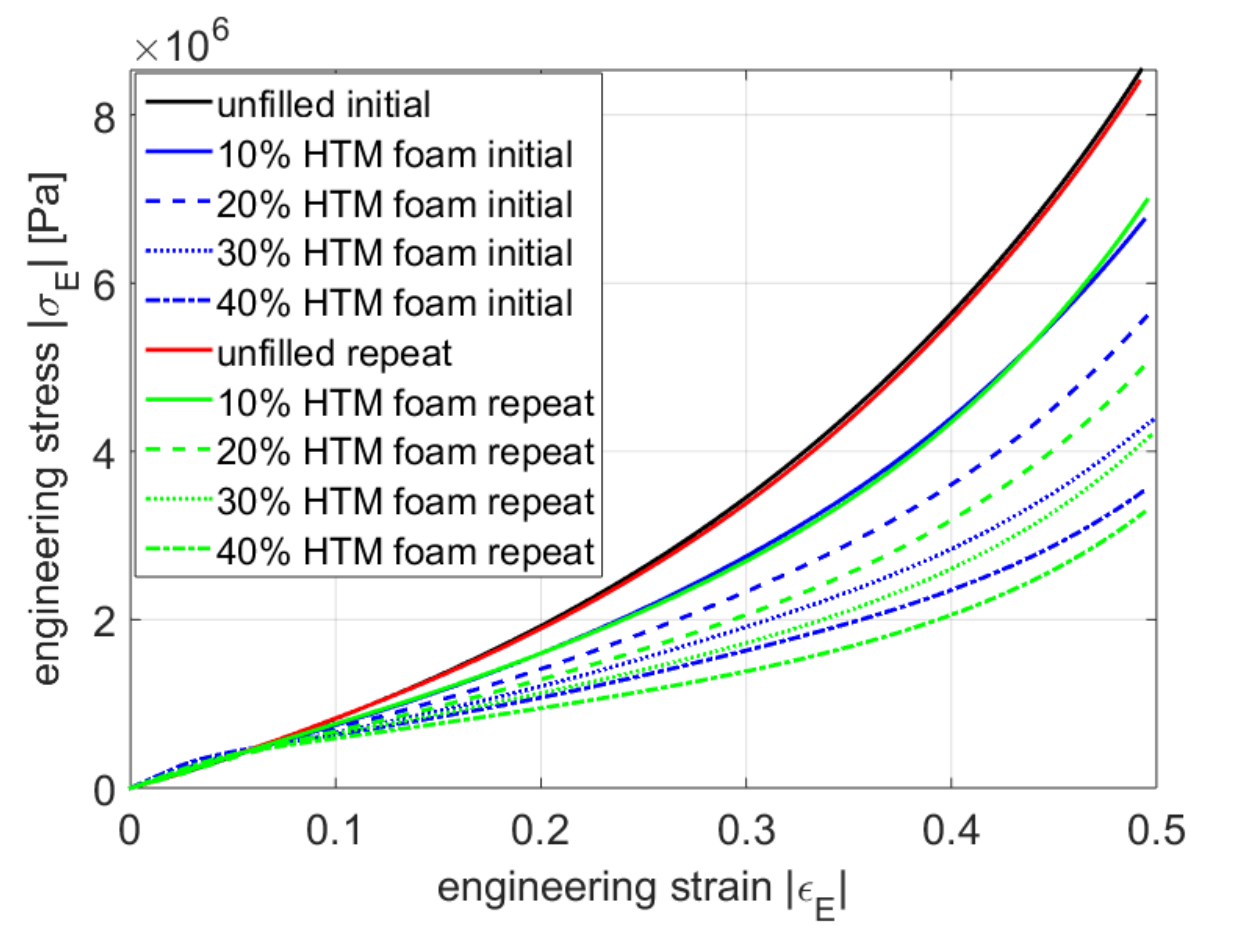}
}

\caption[Stress-strain test data     up to  \subref{fig:92025prec}      $25\%$ and \subref{fig:92050prec}      $50\%$   strain. Results   given for initial      (blue) and repeated  testing  (green).]{(Caption in list of figures) \label{fig:recoverabilityfigs}}

  \end{figure}
\subsection{Stiffness recoverability}
Fig.~\ref{fig:92025prec}   presents test data for the   axial stress-strain curves       to a maximum strain of 25\%   (blue curves), and     stress-strain   results  for the same samples, to the same peak strain level, after 1 week of recovery (green curves). Generally excellent stiffness recoverability (i.e., no permanent softening) was observed for   all  filling fractions $\phi$ to this   strain level, consistent with earlier work by some of the authors \cite{yousaf2020compression}.  Fig.~\ref{fig:92050prec}    shows test data for an identical experiment up to a peak strain  of 50\%,  where     strong stiffness recoverability was observed to high filling fractions. No visible damage was observed in all samples up to 50\% peak strain. For comparison, HGM syntactic foams would typically fracture at such high strain levels \cite{koopman2004compression} leaving them     structurally compromised.

 \section{Mathematical modelling and discussion} 
\label{sec:mathmod}

\subsection{Phenomenological strain energy modelling}  
To describe the mechanical response of HTM syntactic foams at each   filling fraction $\phi$,    a  strain energy function  \cite{ogden1997non,schrodt2005hyperelastic} was sought in the form 
\begin{equation}
\label{eq:genansatz}
W = \Phi(\lambda_1,\lambda_2,\lambda_3) + f(J),
\end{equation}
where $\Phi(\lambda_1,\lambda_2,\lambda_3)$ is a function of principal stretches $\lambda_j$ and $f(J)$ is the {\it compressibility condition}.  Through an appropriate choice of $\Phi(\lambda_1,\lambda_2,\lambda_3)$ and $f(J)$,  an  accurate  recovery of both the axial stress-strain and   transverse strain behaviour of the syntactic foams is possible by differentiating the strain energy function \eqref{eq:genansatz}.  

Following   Figs.~\ref{fig:lzlr920} and \ref{fig:poisson1}, compressible neo-Hookean and compressible Ogden (type-I) strain energy  models \cite{treloar1975physics,ogden1997non}, were considered, as their associated   $f(J)$ forms are generally able to  capture the   transverse   response at dilute filling fractions. For non-dilute    foams,    Ogden (type-II)    models \cite{treloar1975physics,ogden1997non} were investigated.   Results   were obtained using     ABAQUS \cite{ABAQUSm} with tables of coefficients      given in the Supplementary Material. Also presented are figures for the relative errors which are functions of strain   \cite{destrade2017methodical} 
\begin{equation}
\mathrm{relerr}_\sigma(\varepsilon_\mathrm{E}) = \left| \frac{ \sigma_\mathrm{E}^\mathrm{fitted}(\varepsilon_\mathrm{E}) }{ \sigma_\mathrm{E}^\mathrm{test \, data}(\varepsilon_\mathrm{E}) } - 1\right|, 
\qquad \mbox{and}\qquad
\mathrm{relerr}_\lambda(\lambda_\rmz) = \left| \frac{ \lambda_\rmr^\mathrm{fitted}(\lambda_\rmz) }{ \lambda_\rmr^\mathrm{test \, data}(\lambda_\rmz) } - 1\right|,
\end{equation}
with tables of relative errors also given in the Supplementary Material. For a discussion of the limitations and challenges of non-linear least squares fitting to experimental results, see Ogden {\it et.~al.~}\citep{ogden2004fitting}.

\begin{figure}[t]
\centering
\subfigure[  \label{fig:NH_all_920_strstr1} ]{
\includegraphics[width=0.32\textwidth]{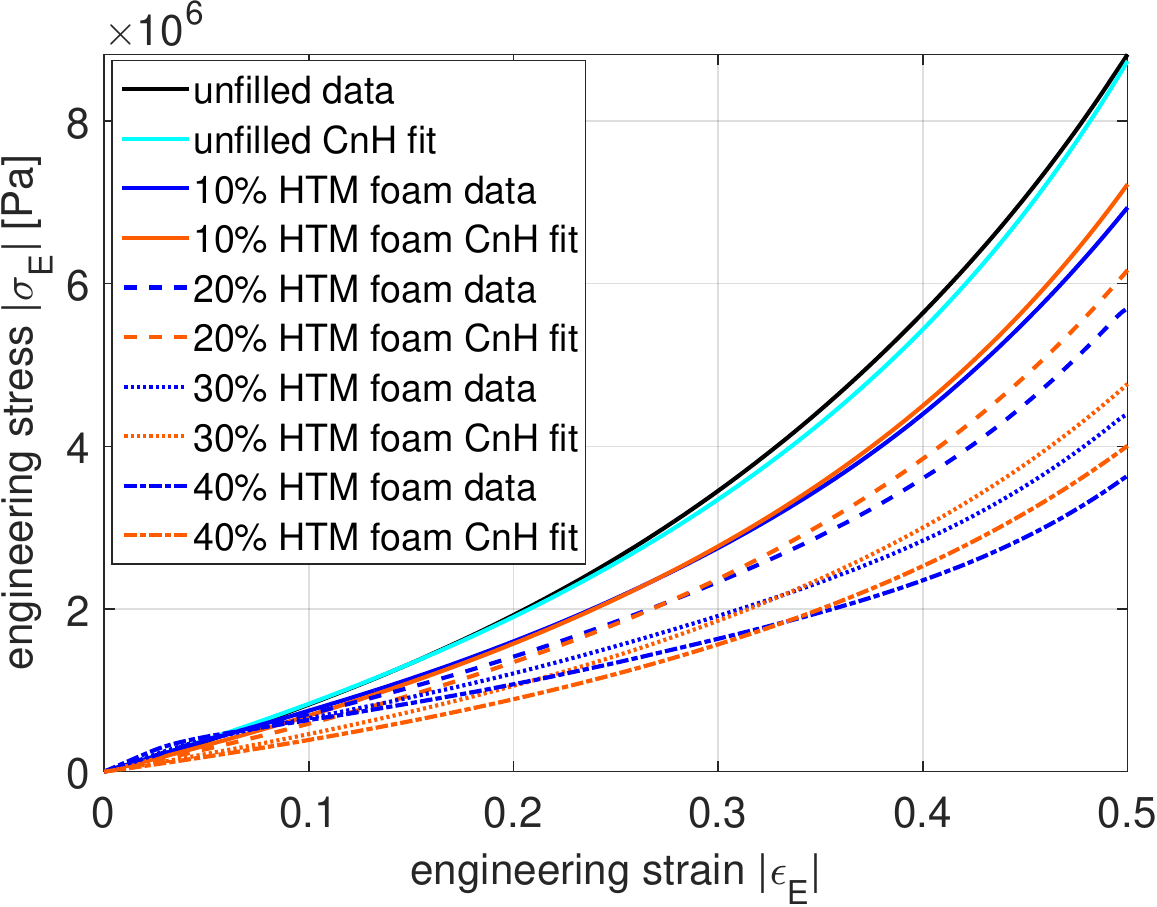}
}
 \subfigure[  \label{fig:NH_all_920_strstr2} ]{
\includegraphics[width=0.32\textwidth]{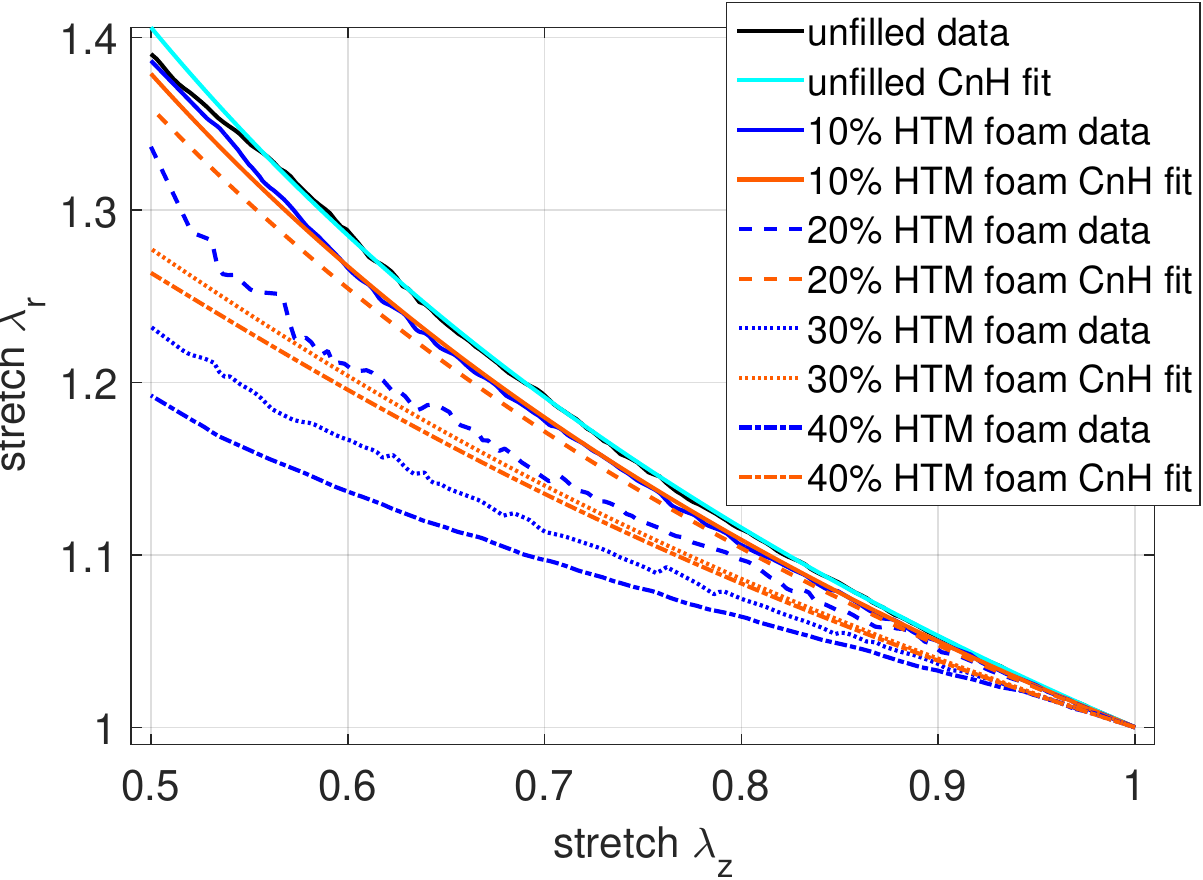}
}
 \subfigure[  \label{fig:NH_all_920_relerror} ]{
\includegraphics[width=0.28\textwidth]{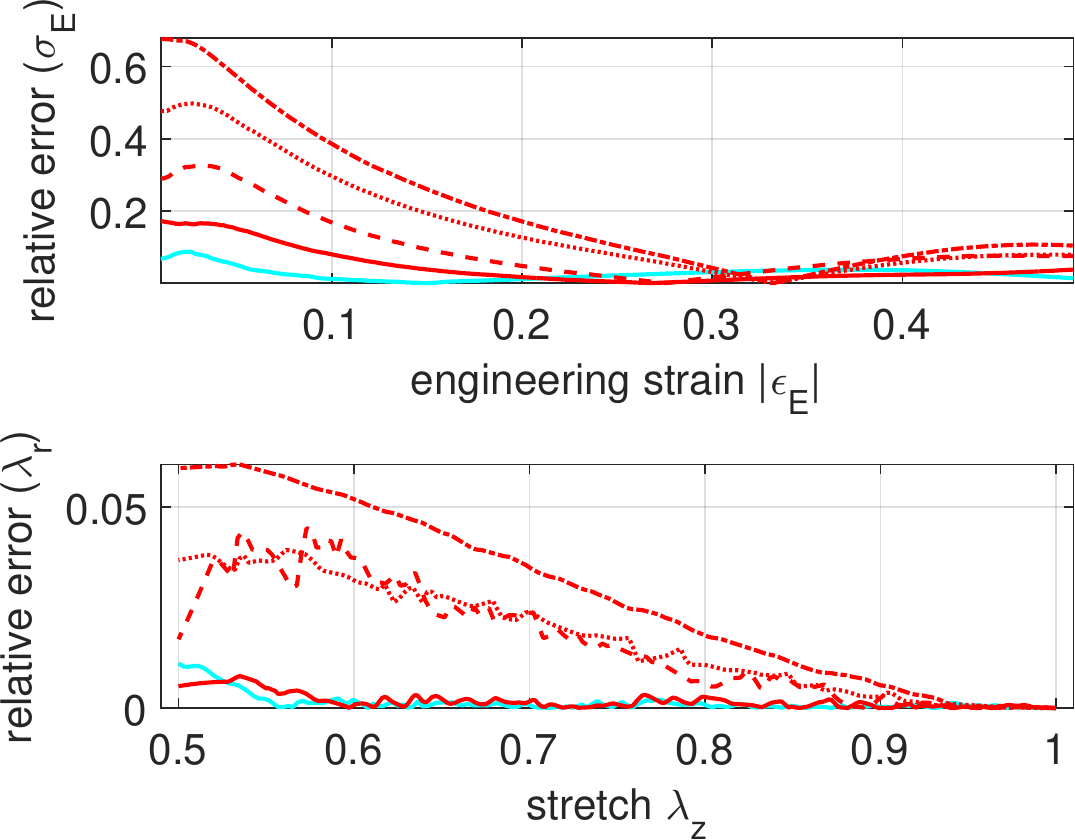}
}
\caption[  Comparison of compressible neo-Hookean fitted models    against test data: \subref{fig:NH_all_920_strstr2}   axial stress-strain  and \subref{fig:NH_all_920_strstr1}     transverse stretch   response.]{(Caption in list of figures) \label{fig:920NHall}}

  \end{figure}
\subsection{Compressible neo-Hookean} \label{sec:cNH}
\begin{subequations}
In this   model, the strain energy takes the form
\begin{equation}
\label{eq:CnHse}
W = C_{10} (\overline{\lambda}_1^2 + \overline{\lambda}_2^2 + \overline{\lambda}_3^2 - 3) +\frac{1}{D_1}(J - 1)^2,
\end{equation}
where 
 $\overline{\lambda}_j = J^{-1/3} \lambda_j$,   $J= \lambda_1 \lambda_2 \lambda_3$, and $C_{10}$ and $D_{1}$ denote real constants. Under uniaxial compression ($\lambda_3 = \lambda_\rmz$, $\lambda_1 = \lambda_2 = \lambda_\rmr$), the engineering stress takes the form
 \begin{equation}
 \sigma_\mathrm{E} = \frac{\partial W}{\partial \lambda_\rmz} = \frac{4C_{10} J^{-2/3}}{3 \lambda_\rmz} \left( \lambda_\rmz^2 - \frac{J}{\lambda_\rmz} \right) + \frac{2J}{D_1\lambda_\rmz} (J - 1),
 \end{equation}
 where $J$ is obtained by solving the {\it compressibility relation}, given by imposing zero transverse stress  conditions (i.e.,   $\sigma_1  = 0$) and takes the form
 \begin{equation}
 \label{eq:cccnh}
 \frac{2C_{10} J^{-5/3}}{3} \left[ \frac{J}{\lambda_\rmz} - \lambda_\rmz^2 \right] + \frac{2}{D_1} (J - 1) = 0.
 \end{equation}
  \end{subequations}

Fig.~\ref{fig:920NHall} presents   the stress-strain and transverse stretch response for   CnH fitted models   \eqref{eq:CnHse}, for all   filling fractions (test data   superposed in blue). These figures reveal that this strain energy form   only     qualitatively describes    dilute syntactic foams, with significant relative errors for the stress (relerr$_\sigma \gg 10\%$), except for the unfilled polyurethane. Generally, the relative error for the transverse response was well-controlled. Given the poor performance of this strain energy ansatz we now proceed to more advanced models.

\begin{figure}[t]

\centering

\subfigure[  \label{fig:ogdenI_92000_stress} ]{
\includegraphics[width=0.30\textwidth]{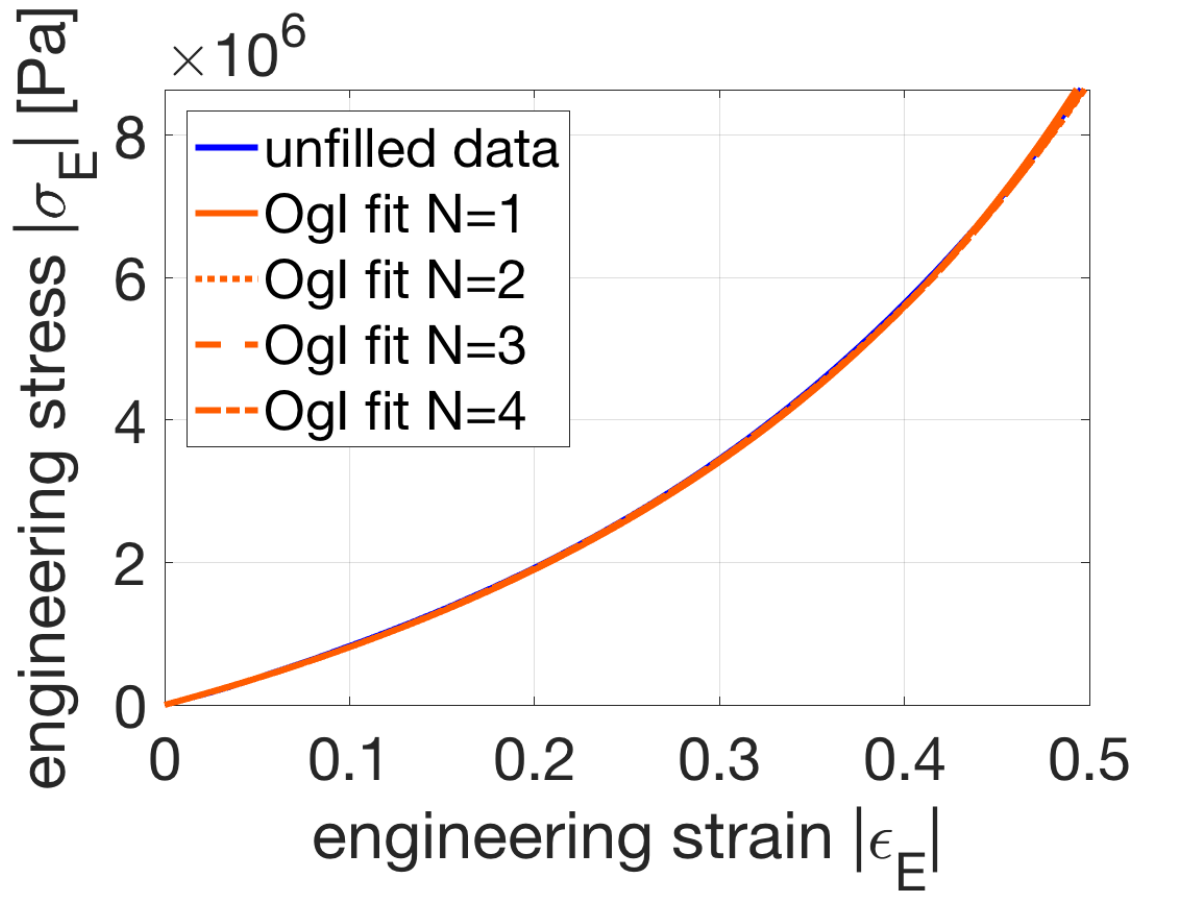}
}
 \subfigure[  \label{fig:ogdenI_92000_tstrain} ]{
\includegraphics[width=0.30\textwidth]{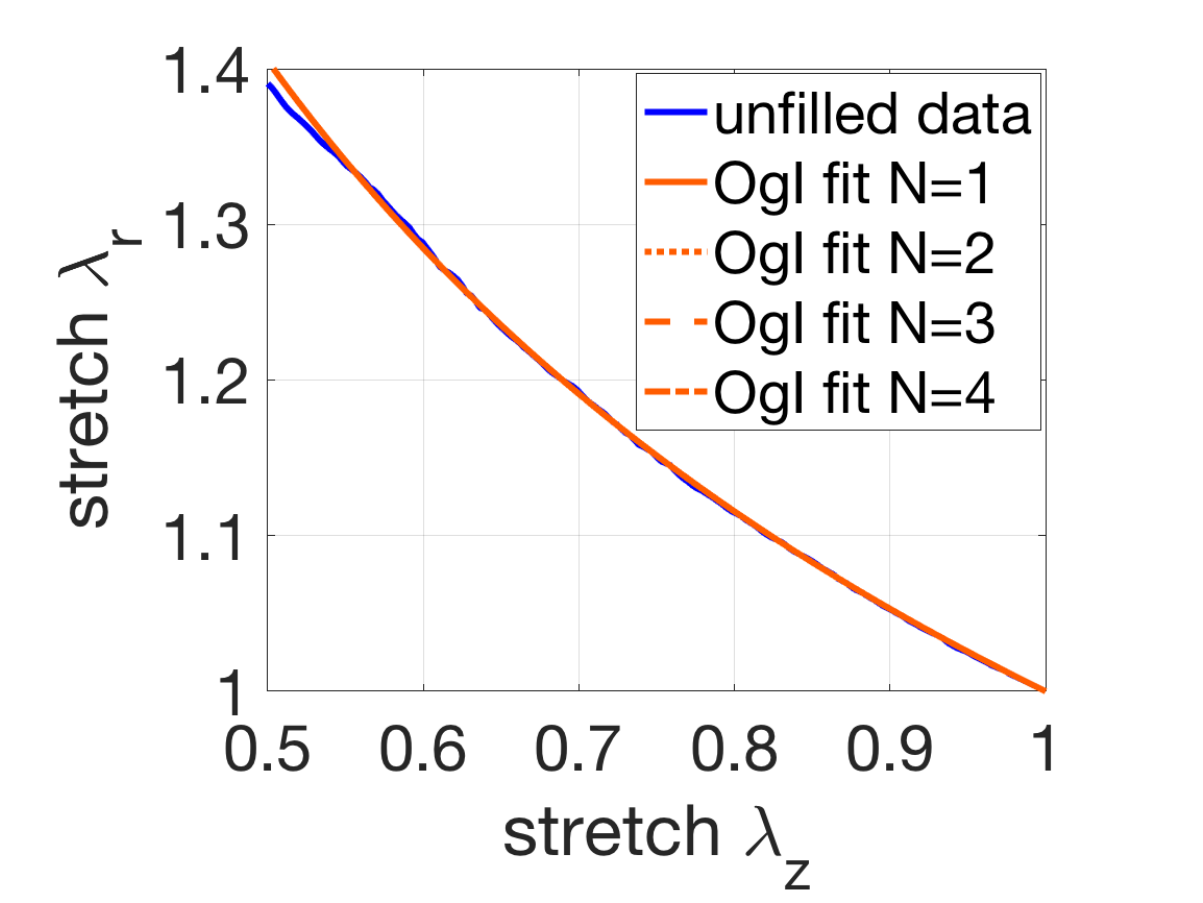}
}
 \subfigure[  \label{fig:og00_all_920_relerror} ]{
\includegraphics[width=0.28\textwidth]{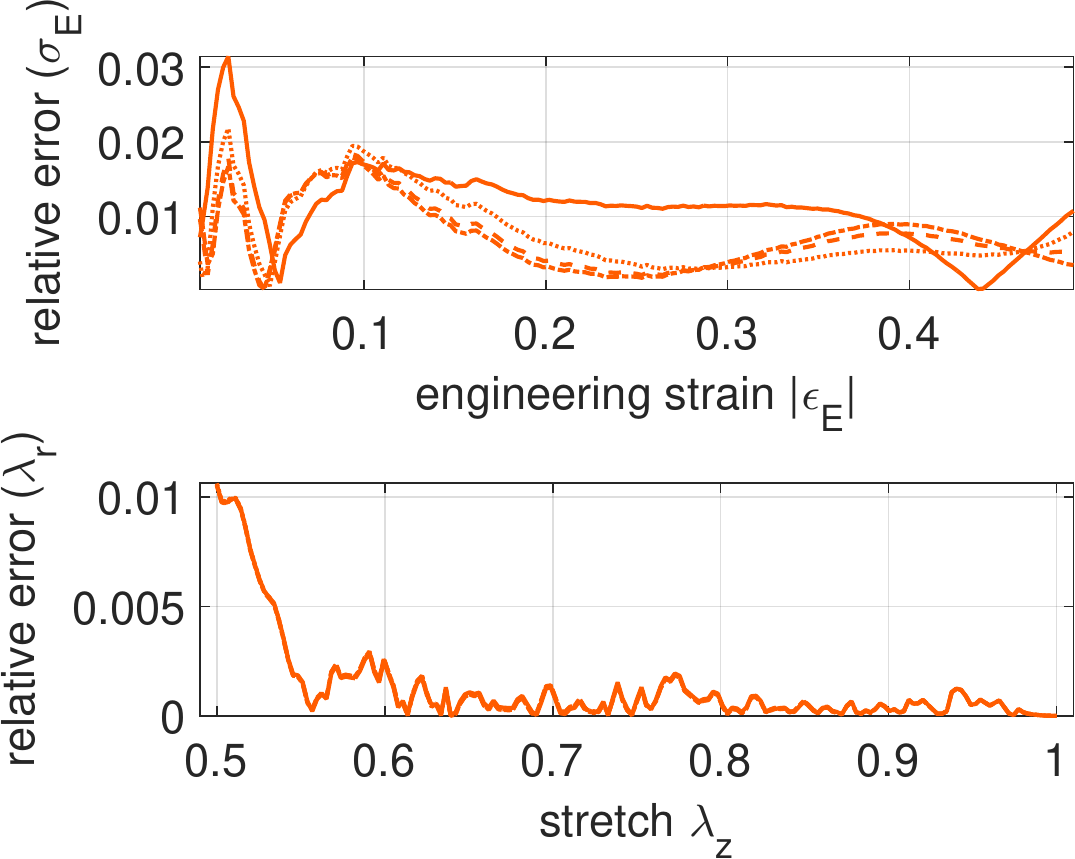}
}
\\
 \subfigure[  \label{fig:ogdenI_92010_stress} ]{
\includegraphics[width=0.30\textwidth]{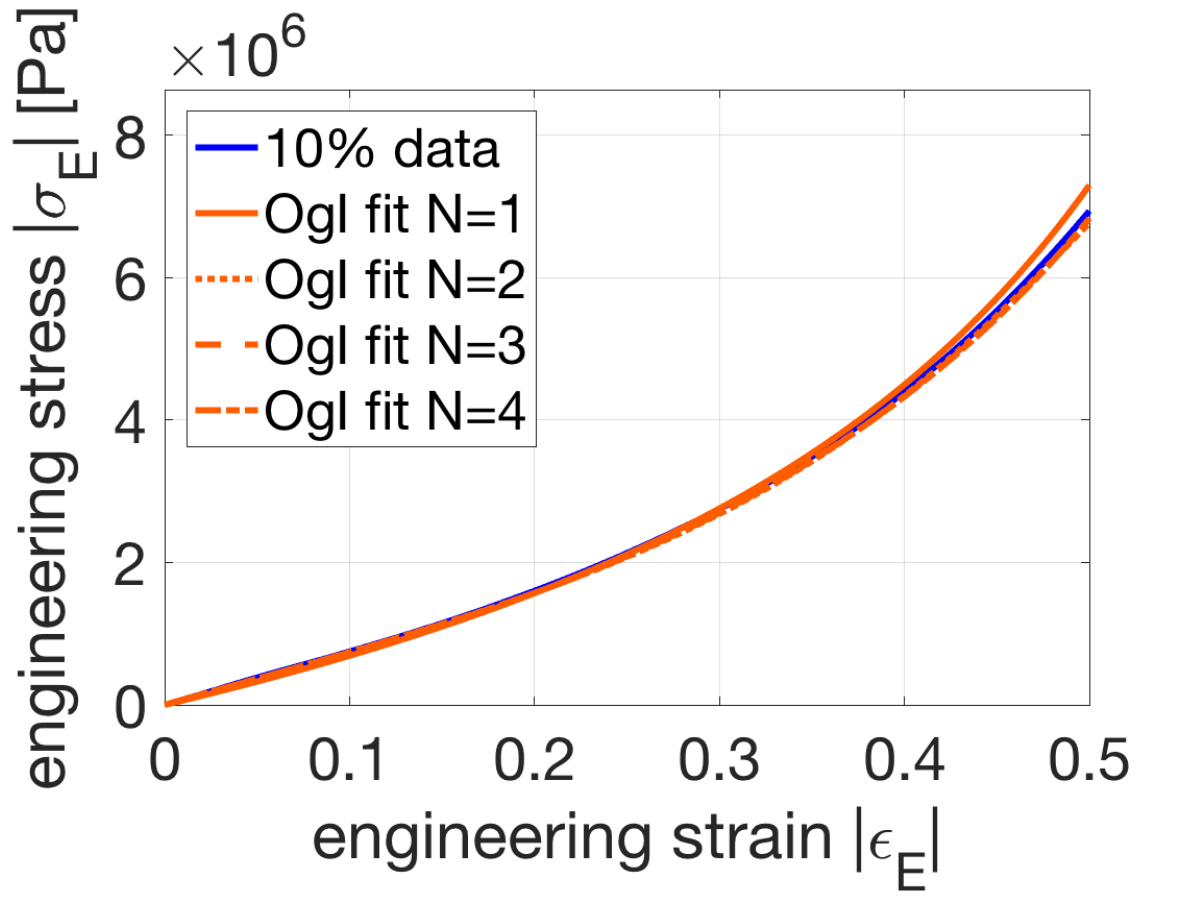}
}
 \subfigure[  \label{fig:ogdenI_92010_tstrain} ]{
\includegraphics[width=0.30\textwidth]{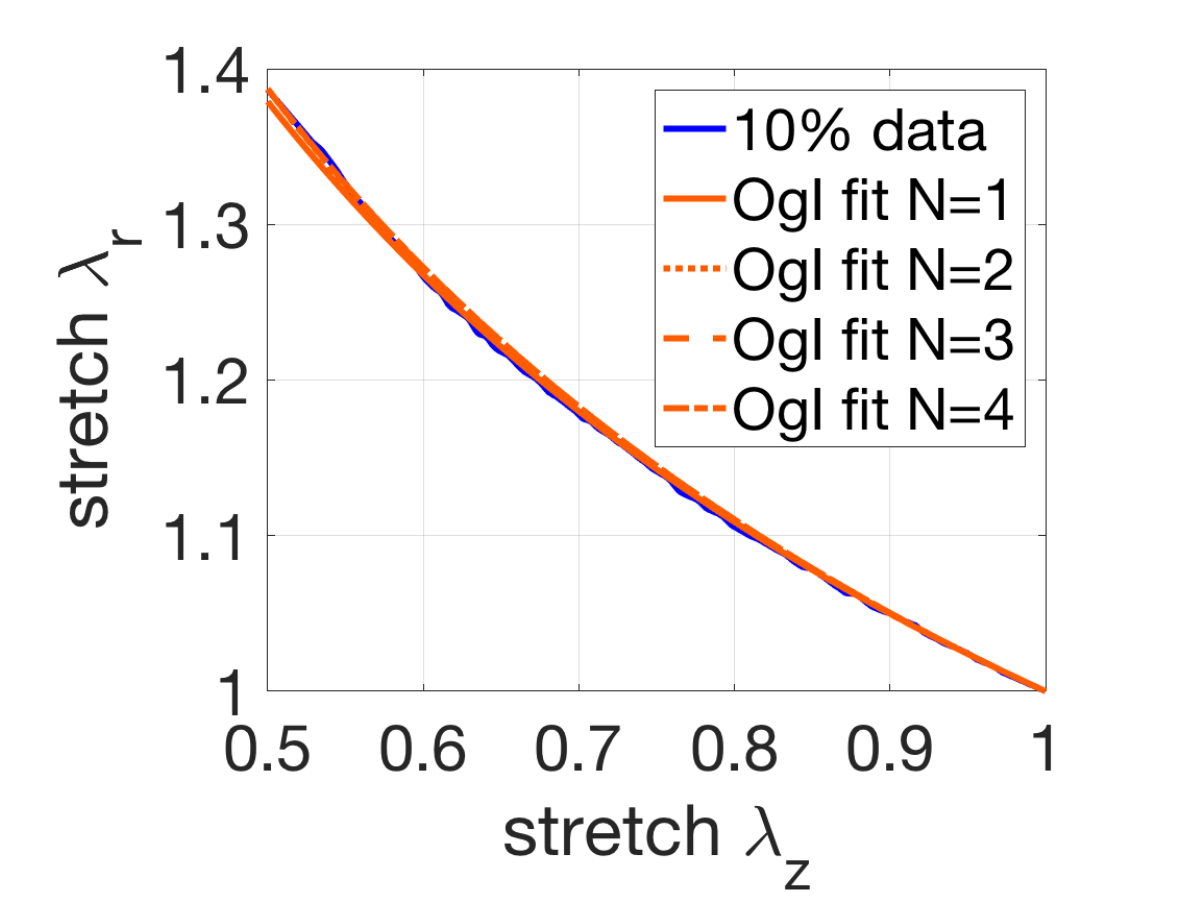}
}
  \subfigure[  \label{fig:og10_all_920_relerror} ]{
\includegraphics[width=0.28\textwidth]{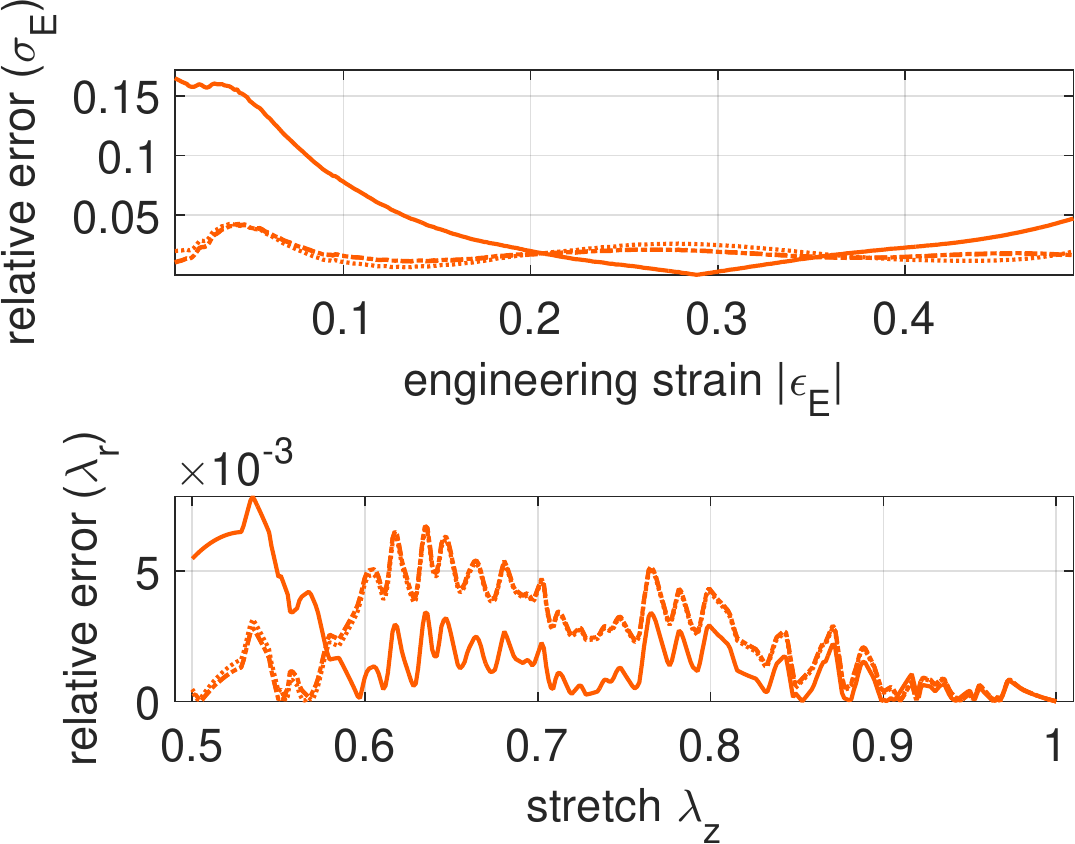}
}

\caption[Comparison of  Ogden Type I  fitted models   against test data for dilute filling fractions]{(Caption in list of figures)     \label{fig:ogdenI_920allpt1}}

\end{figure}

\begin{figure}[t]

\centering

 \subfigure[  \label{fig:ogdenI_92020_stress} ]{
\includegraphics[width=0.30\textwidth]{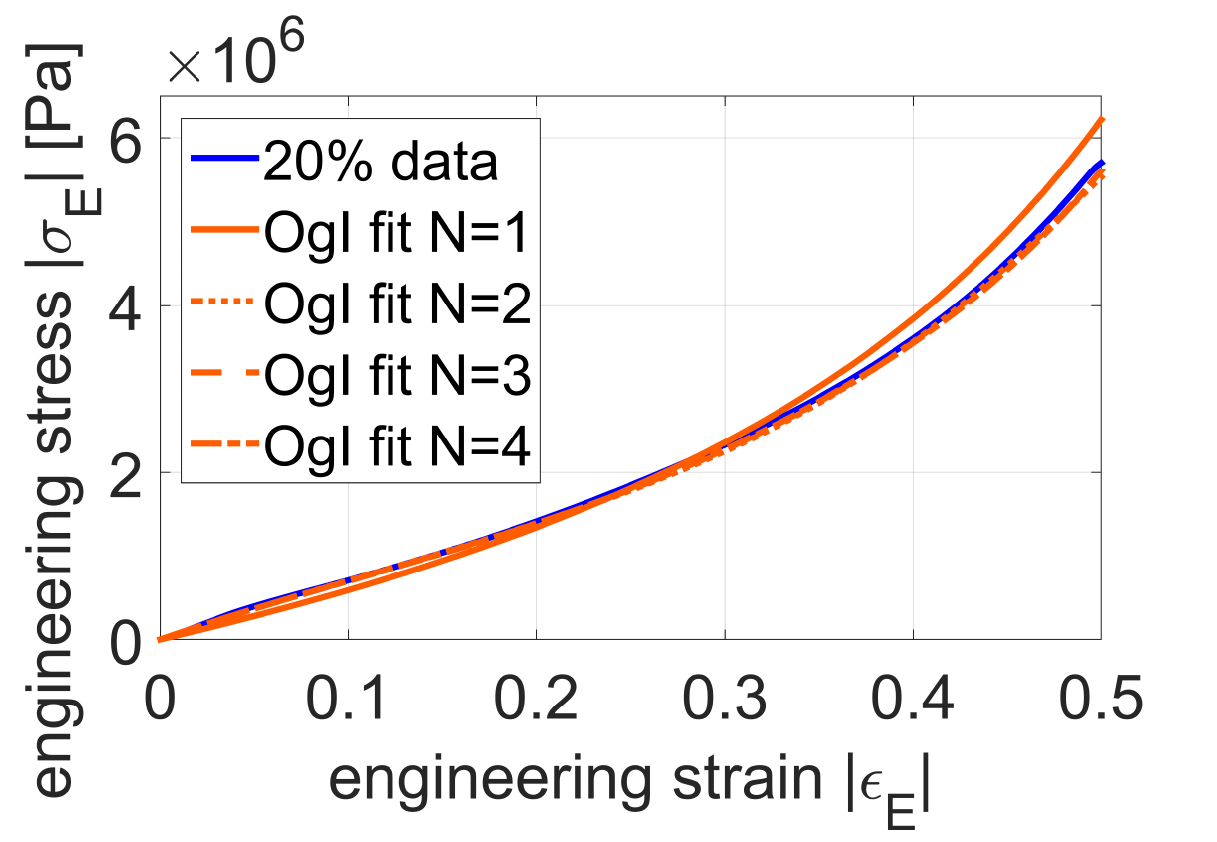}
}
 \subfigure[  \label{fig:ogdenI_92020_tstrain} ]{
\includegraphics[width=0.30\textwidth]{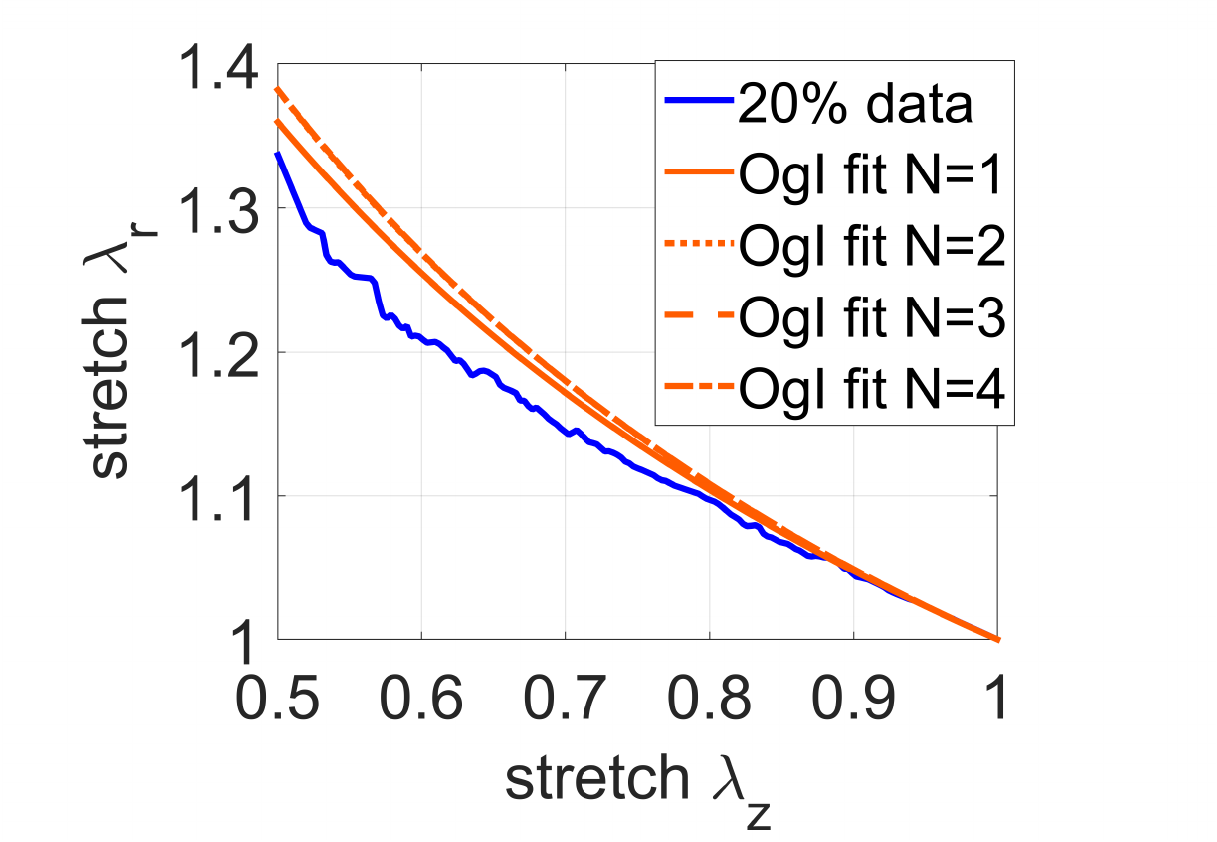}
}
 \subfigure[  \label{fig:ogdenI_all_92020_relerror} ]{
\includegraphics[width=0.28\textwidth]{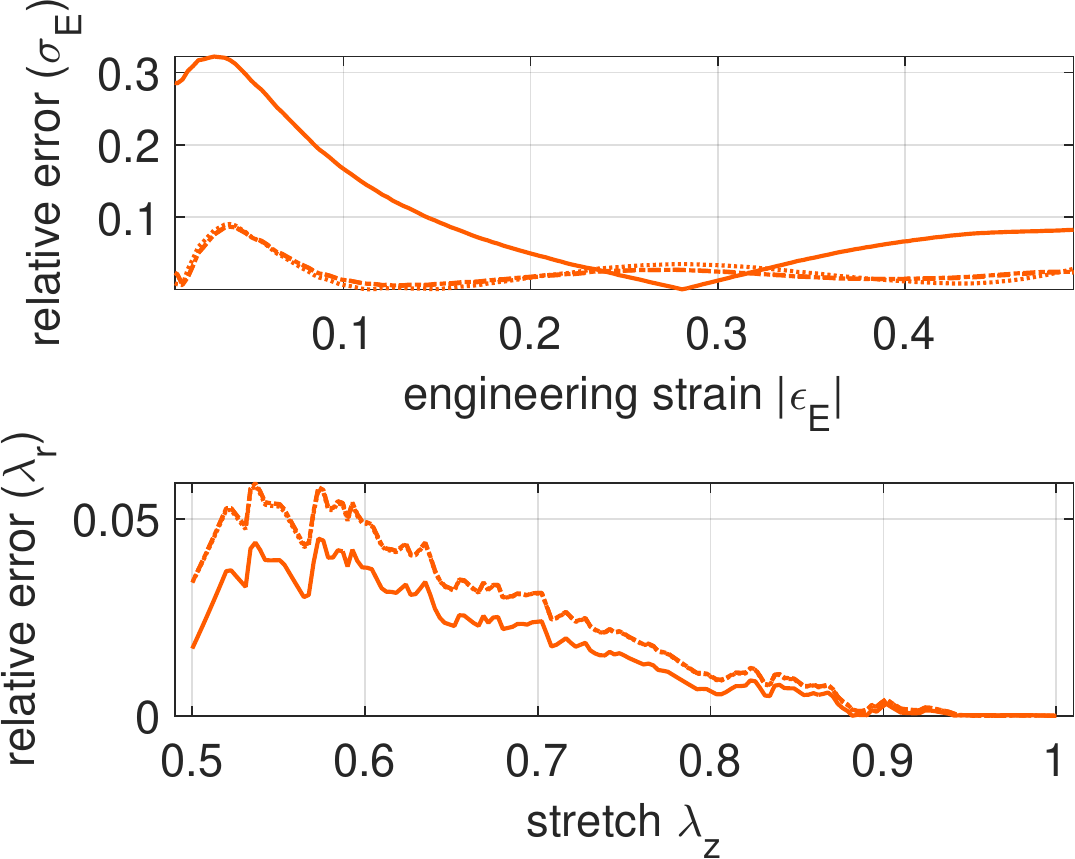}
}
\\
\subfigure[  \label{fig:ogdenI_92030_stress} ]{
\includegraphics[width=0.30\textwidth]{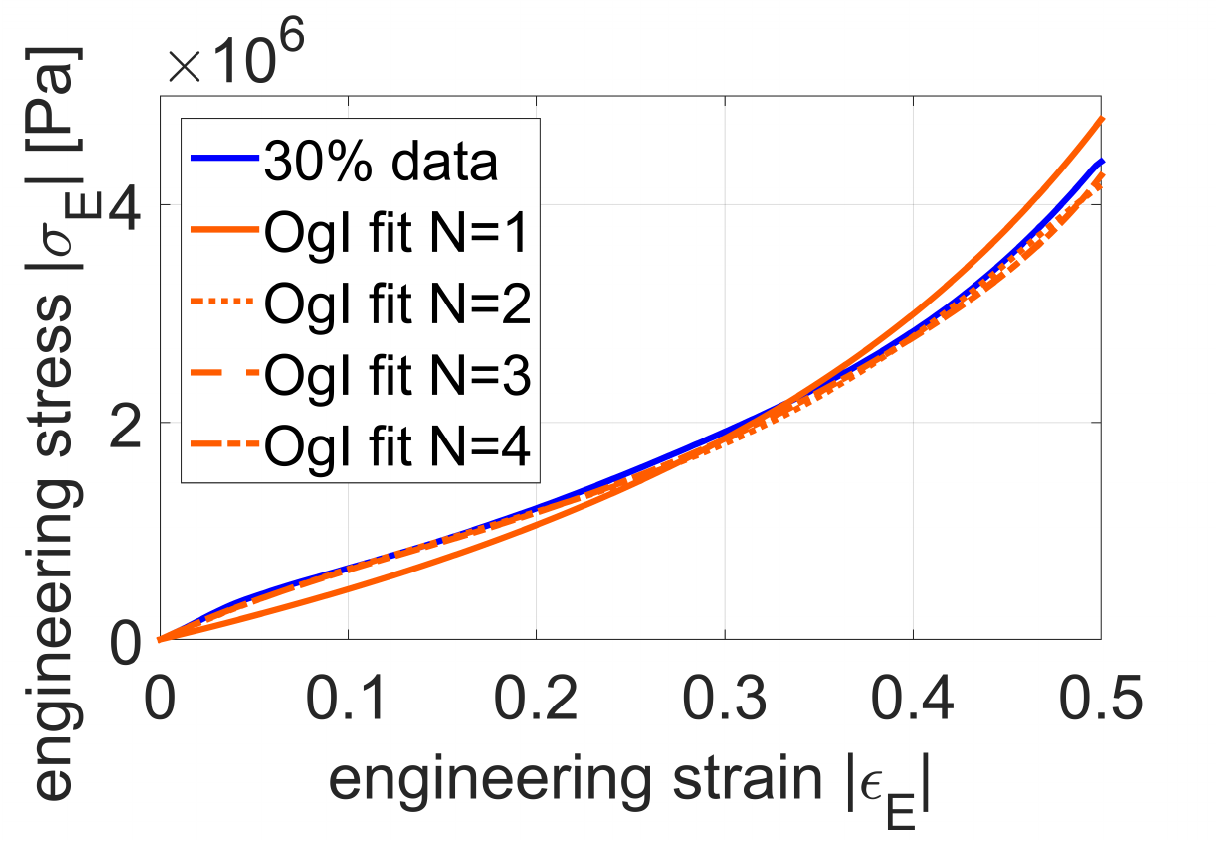}
}
 \subfigure[  \label{fig:ogdenI_92030_tstrain} ]{
\includegraphics[width=0.30\textwidth]{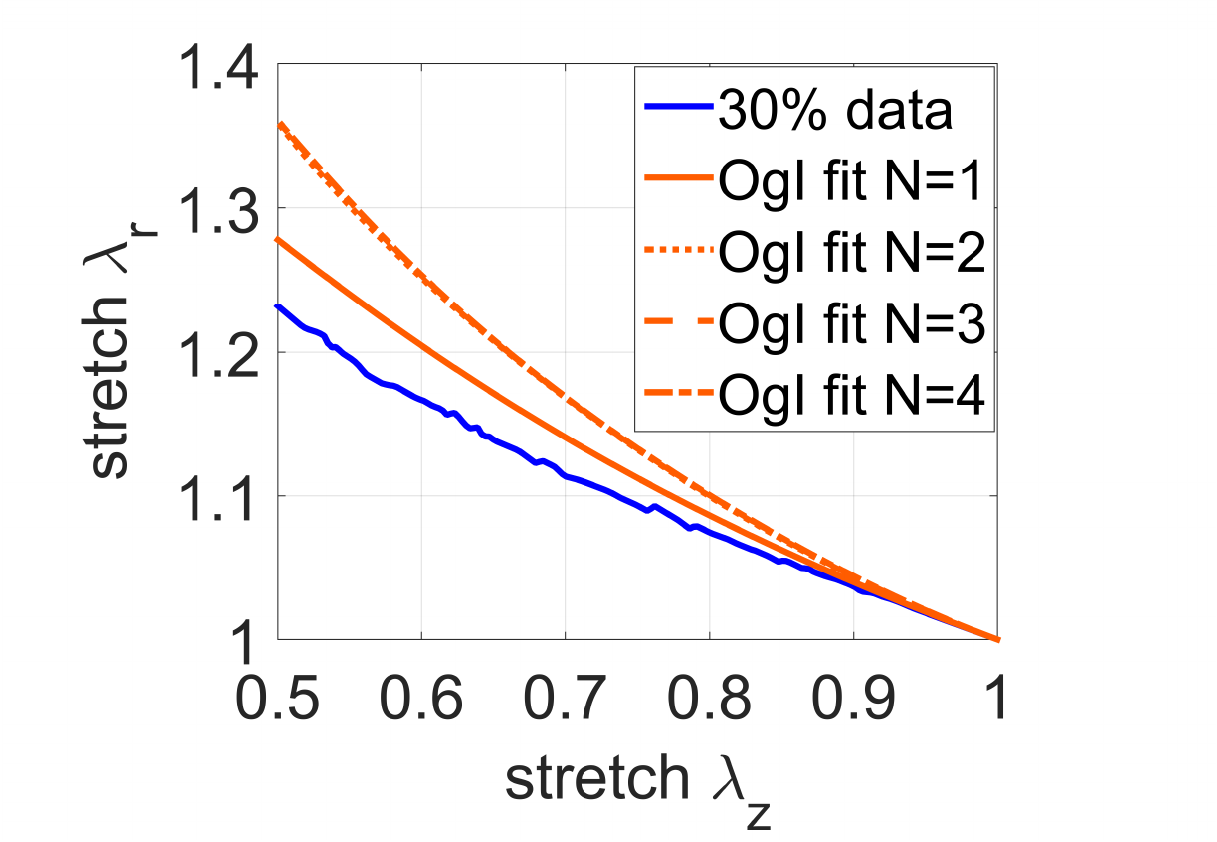}
}
 \subfigure[  \label{fig:ogdenI_all_92030_relerror} ]{
\includegraphics[width=0.28\textwidth]{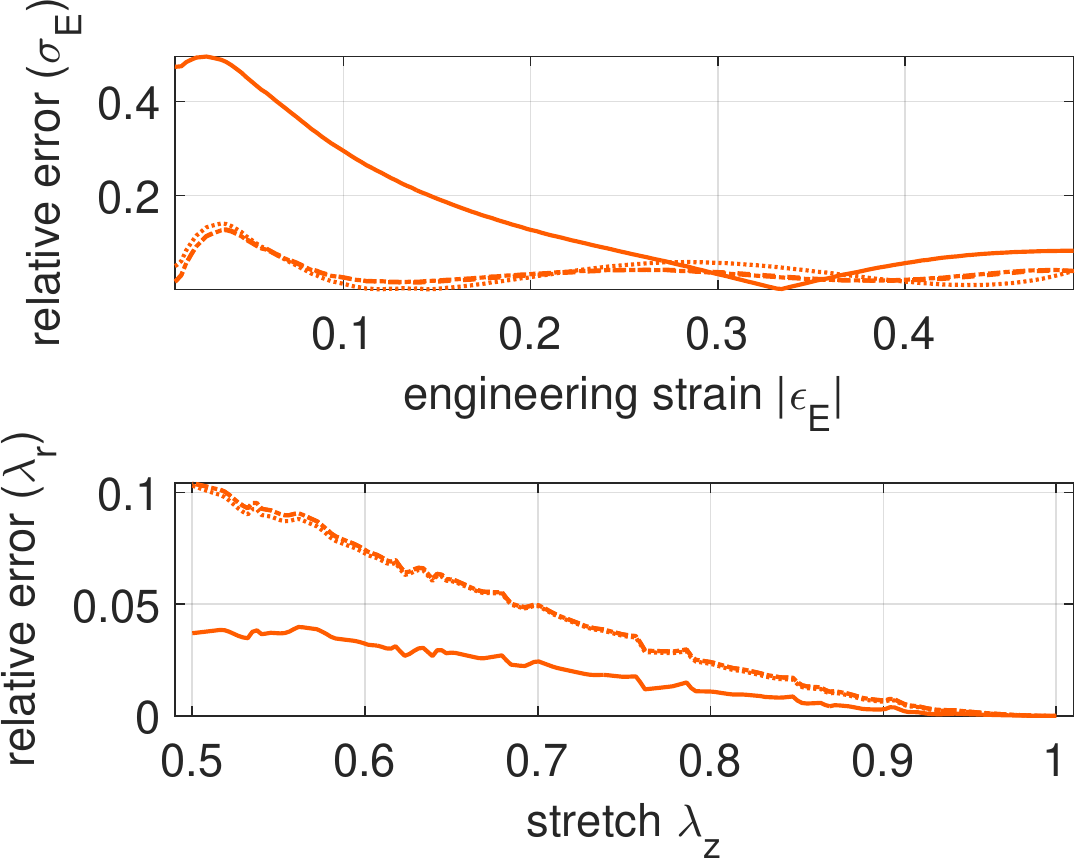}
}
 \\
 \subfigure[  \label{fig:ogdenI_92040_stress} ]{
\includegraphics[width=0.30\textwidth]{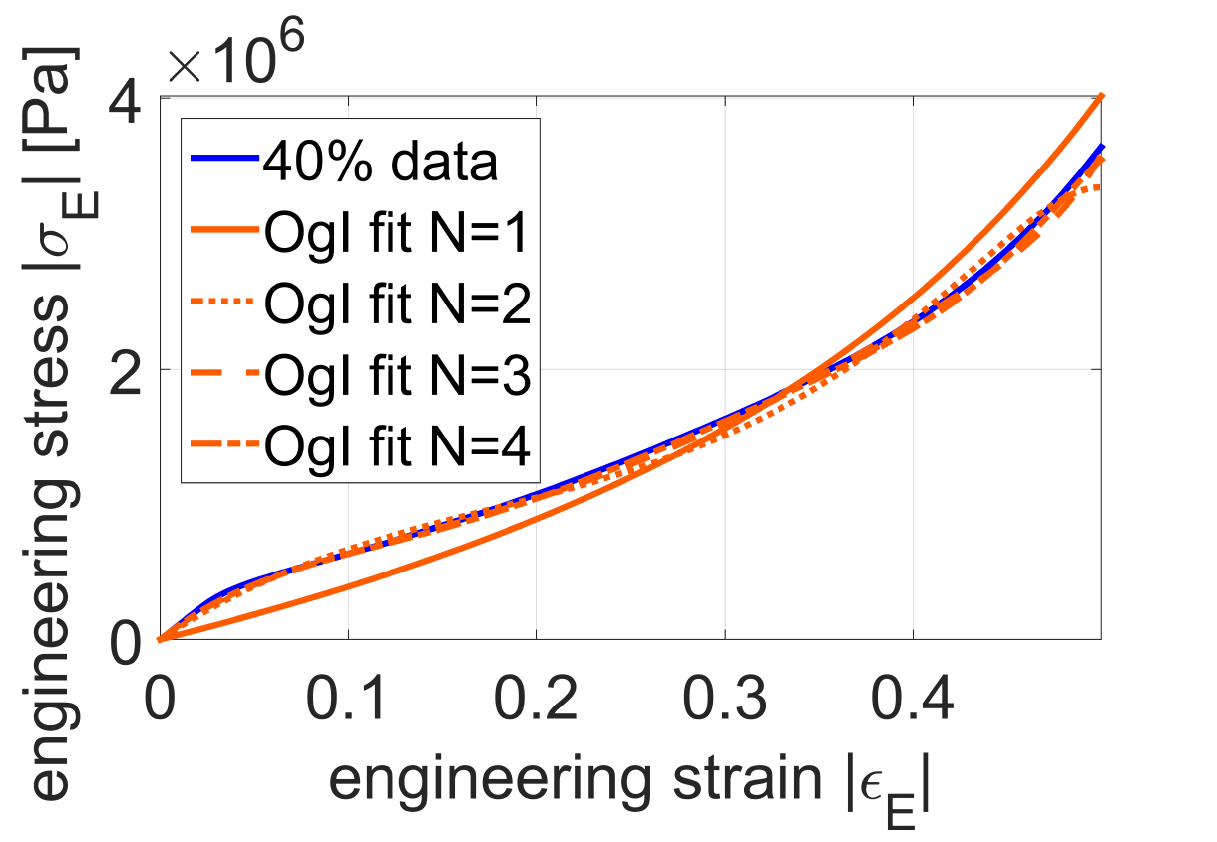}
}
 \subfigure[  \label{fig:ogdenI_92040_tstrain} ]{
\includegraphics[width=0.30\textwidth]{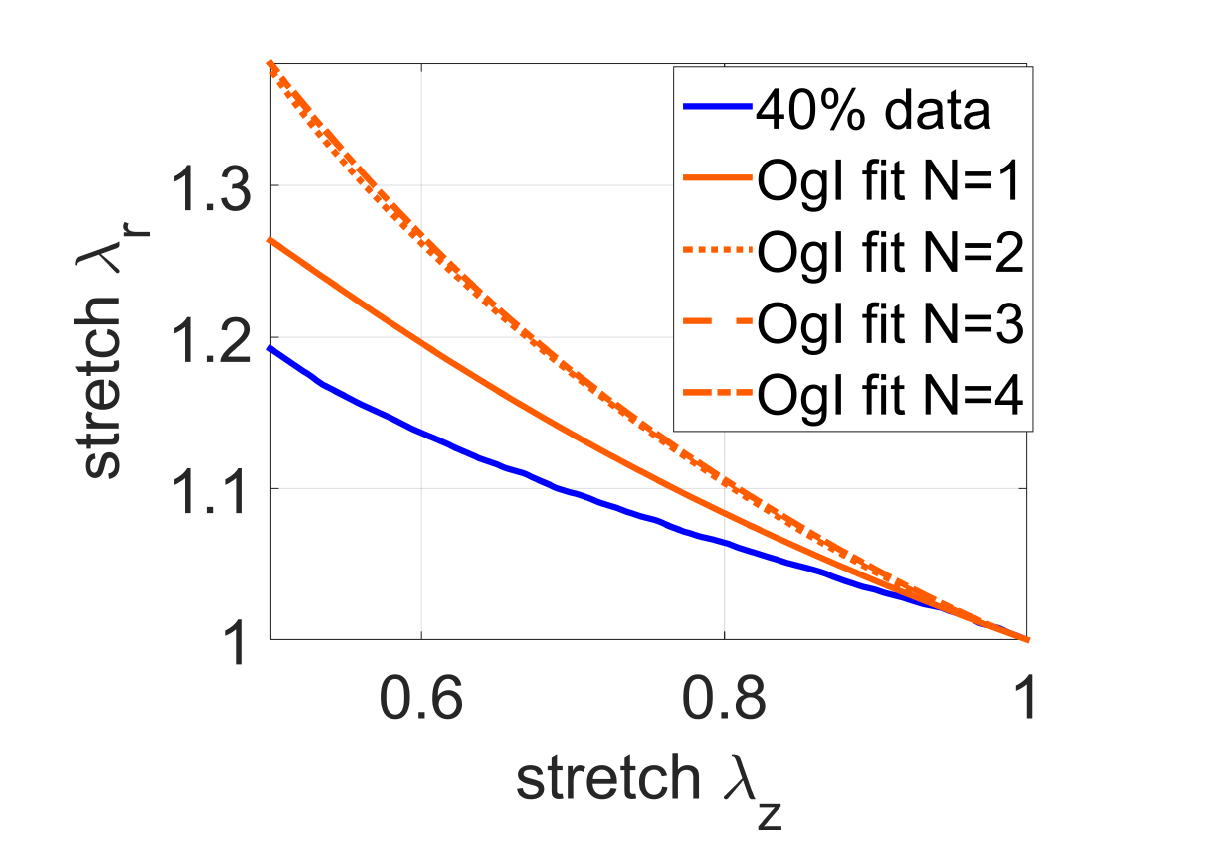}
}
 \subfigure[  \label{fig:ogdenI_all_92040_relerror} ]{
\includegraphics[width=0.28\textwidth]{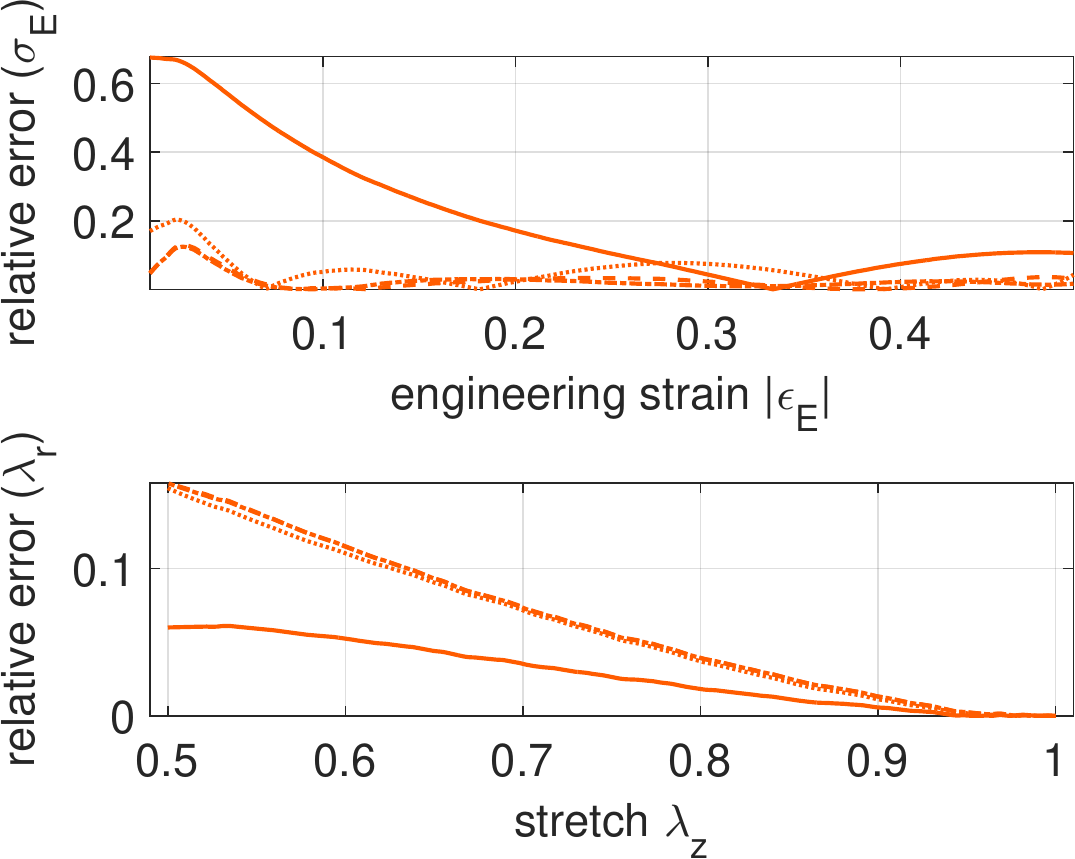}
}
\caption[Comparison of  Ogden Type I  fitted models   against test data for larger filling fractions]{(Caption in list of figures)     \label{fig:ogdenI_920allpt2}}

\end{figure}

      \subsection{Compressible Ogden type-I}
      \begin{subequations}
In this model, the strain energy ansatz takes the form
\begin{equation}
\label{eq:cot1}
W = \sum_{j = 1}^{N} \left\{ \frac{2 \mu_j}{\alpha_j^2} \left[ \overline{\lambda}_1^{\alpha_j} + \overline{\lambda}_2^{\alpha_j} + \overline{\lambda}_3^{\alpha_j} - 3 \right] + \frac{1}{D_j} (J - 1)^{2j}\right\},
\end{equation}
      where $\mu_j$, $\alpha_j$, and $D_j$ are real-valued   constants. Under uniaxial compression      
\begin{equation}
\sigma_\mathrm{E} =  \frac{\partial W}{\partial \lambda_\rmz} =  \sum_{k = 1}^{N} \left\{ \frac{4\mu_k}{3 \alpha_k \lambda_\rmz }\left[ \left( J^{-1/3}\lambda_\rmz \right)^{\alpha_k} - \left( J^{-1/3}\lambda_\rmz \right)^{-\alpha_k/2}\right] + \frac{2k J}{D_k \lambda_\rmz} (J - 1)^{2k - 1} \right\},
\end{equation}
where $J$ is obtained by solving the compressibility relation
\begin{equation}
\label{eq:ccogi}
\sum_{k = 1}^{N} \left\{\frac{2 \mu_k}{3 \alpha_k J} \left[  (J^{-1/3}\lambda_\rmz)^{-\alpha_k/2} - (J^{-1/3}\lambda_\rmz)^{\alpha_k} \right] + \frac{2 k}{D_k}(J-1)^{2k - 1} \right\} = 0.
\end{equation}
        \end{subequations}
        Figs.~\ref{fig:ogdenI_920allpt1} and \ref{fig:ogdenI_920allpt2} present  results for Compressible Ogden type-I (OgI) fitted models   for all filling fractions and truncation values   $N$.    In these instances, the axial stress-strain and transverse strain data were  qualitatively well-modelled, even in the simplest setting of $N=1$. Also in these figures we include the relative error for the stress and transverse stretch, which is generally well controlled   in Fig.~\ref{fig:ogdenI_920allpt1}.   However, Fig.~\ref{fig:ogdenI_920allpt2} reveals that   OgI strain energy models do not possess an appropriate $f(J)$    to accurately recover the transverse strain response for $\phi > 10\%$. In Table \ref{tab:bestfits} we present relerr$_\sigma$ and relerr$_\lambda$ corresponding to the best-fitted models for all samples, showing that the Ogden type-I model works well up to filling fractions of $\phi = 20\%$ to within a maximal error deviation of $8.6\%$.

\begin{figure}[t]
\centering

\subfigure[  \label{fig:ogdenII_92020_stress} ]{
\includegraphics[width=0.30\textwidth]{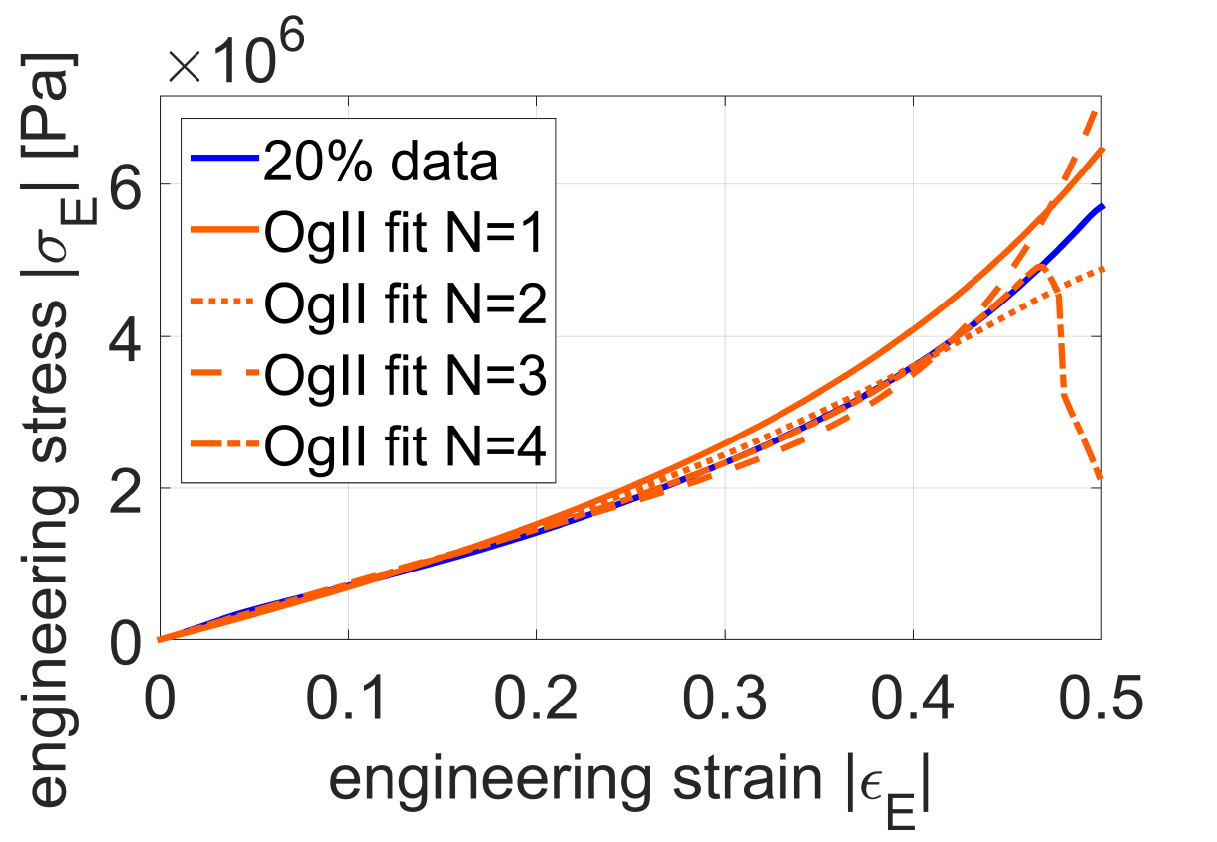}
}
 \subfigure[  \label{fig:ogdenII_92020_tstrain} ]{
\includegraphics[width=0.30\textwidth]{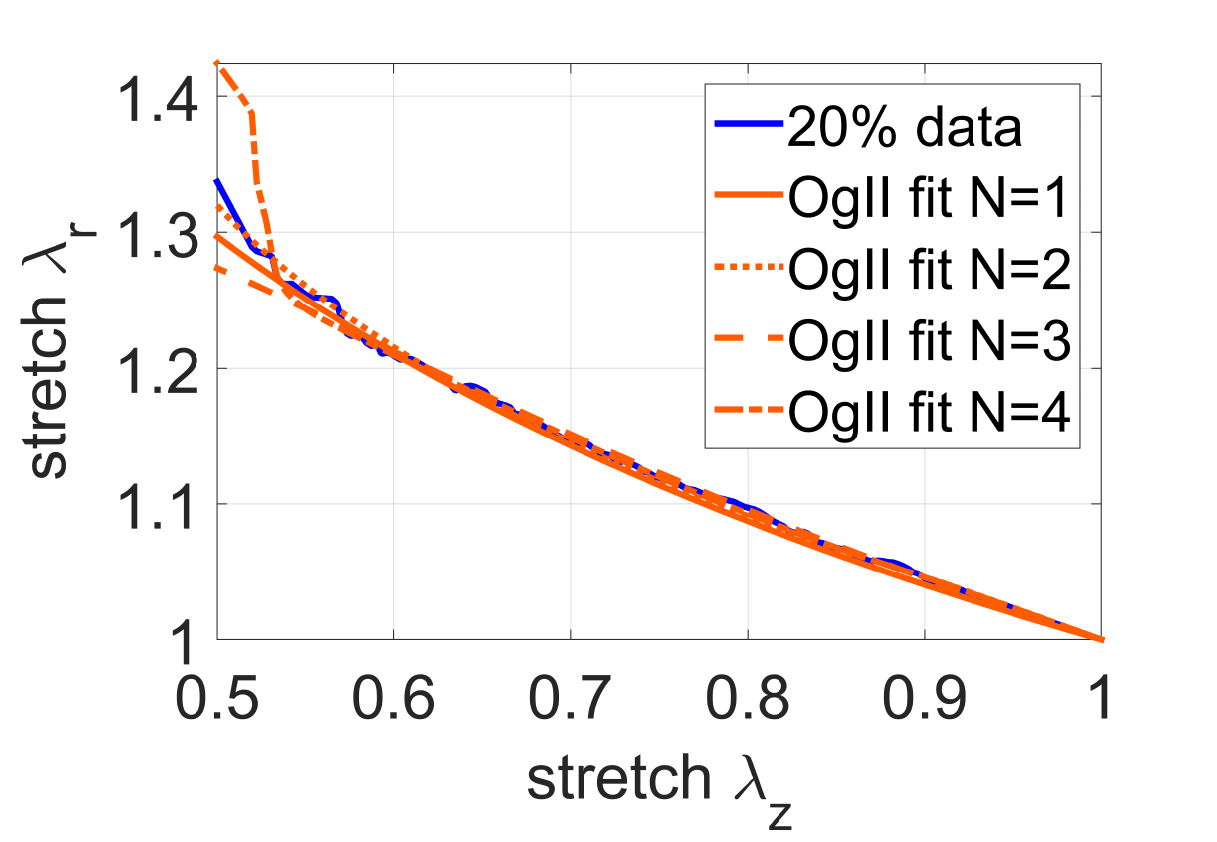}
}
 \subfigure[  \label{fig:ogdenII_92020_relerror} ]{
\includegraphics[width=0.28\textwidth]{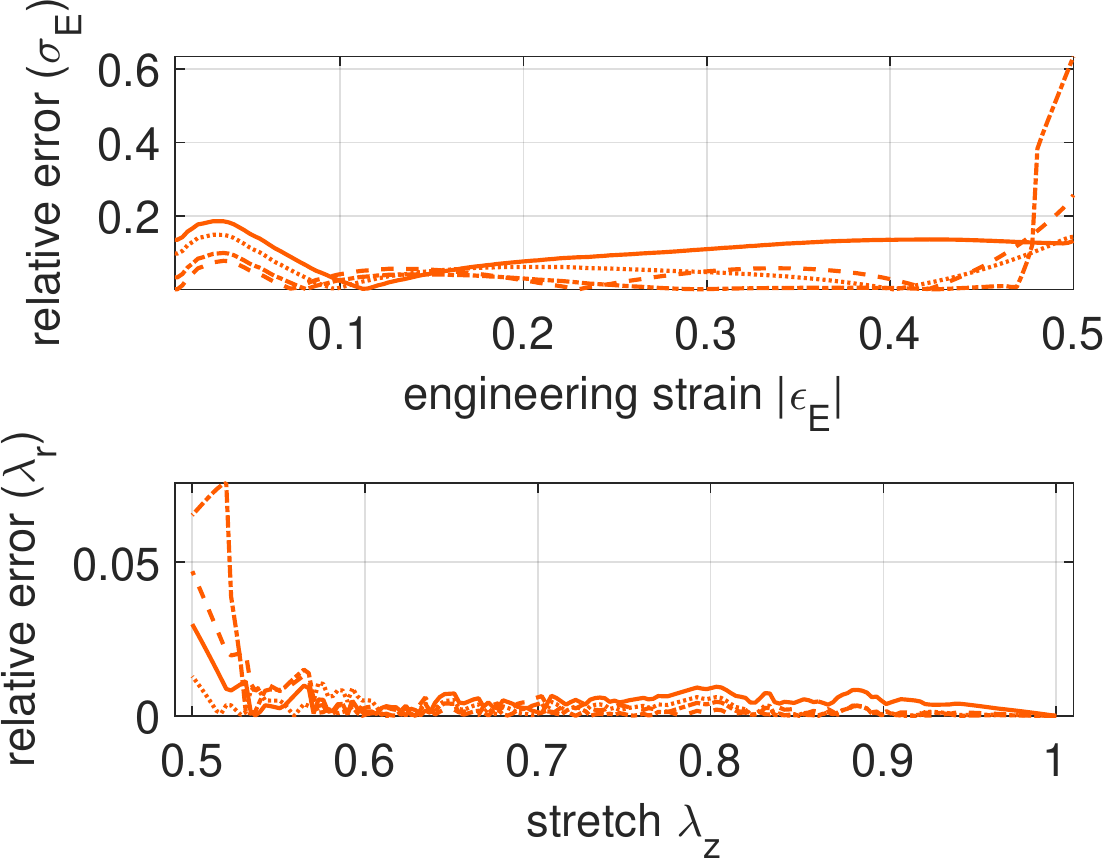}
}
\\
\subfigure[  \label{fig:ogdenII_92030_stress} ]{
\includegraphics[width=0.30\textwidth]{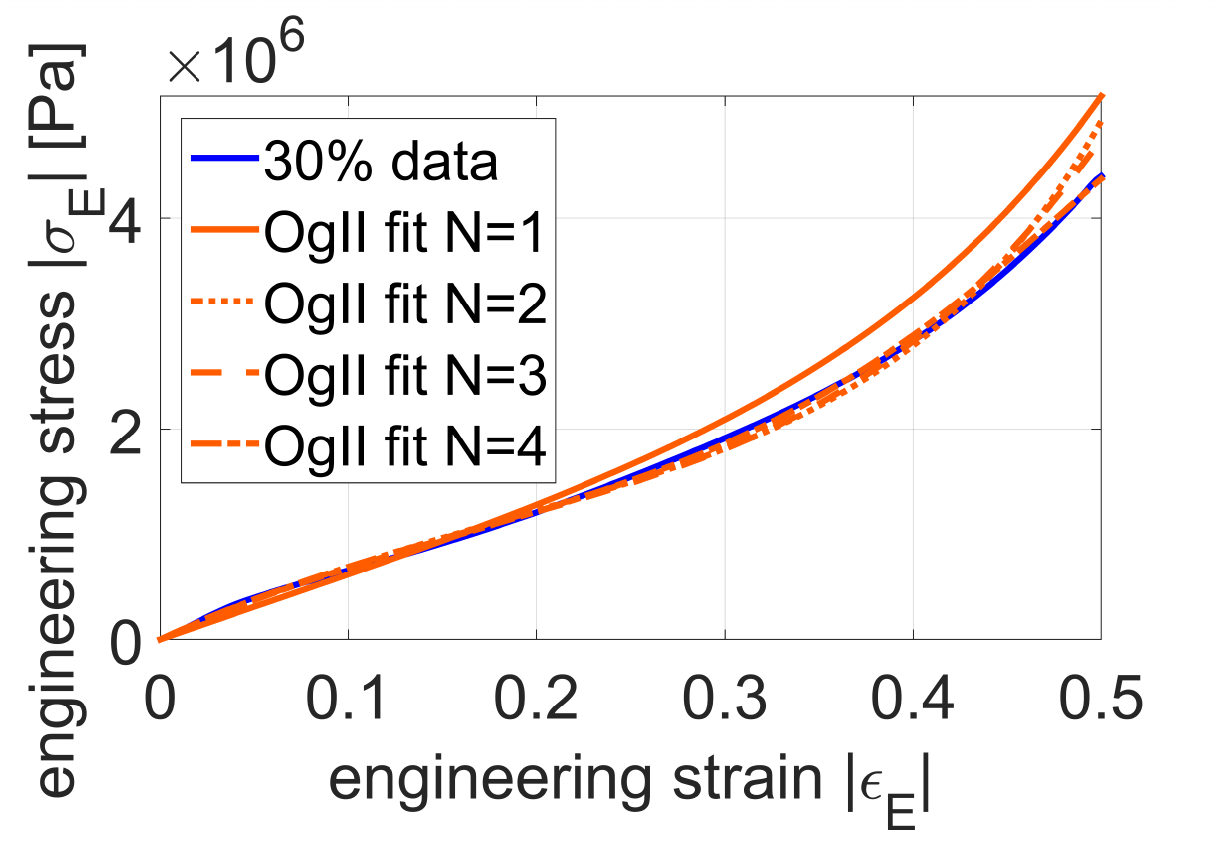}
}
 \subfigure[  \label{fig:ogdenII_92030_tstrain} ]{
\includegraphics[width=0.30\textwidth]{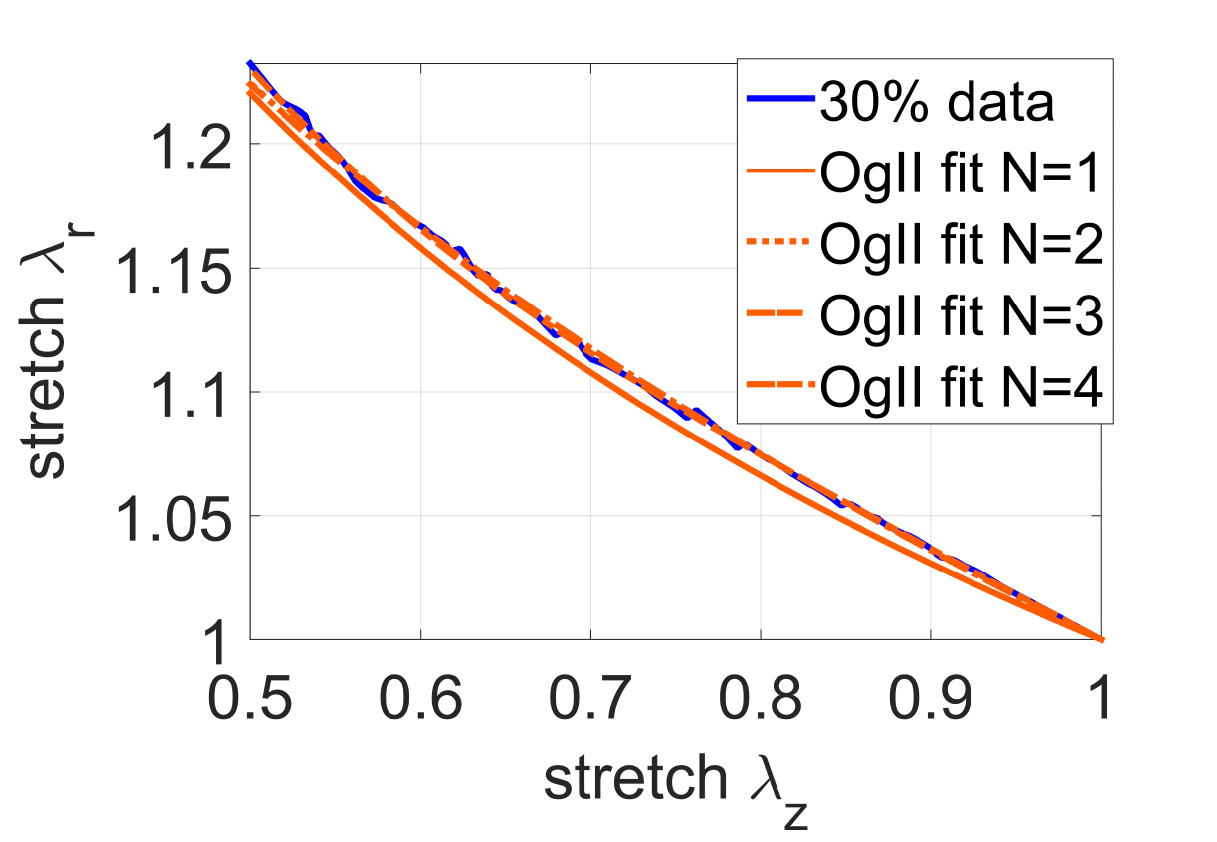}
}
 \subfigure[  \label{fig:ogdenII_92030_relerror} ]{
\includegraphics[width=0.28\textwidth]{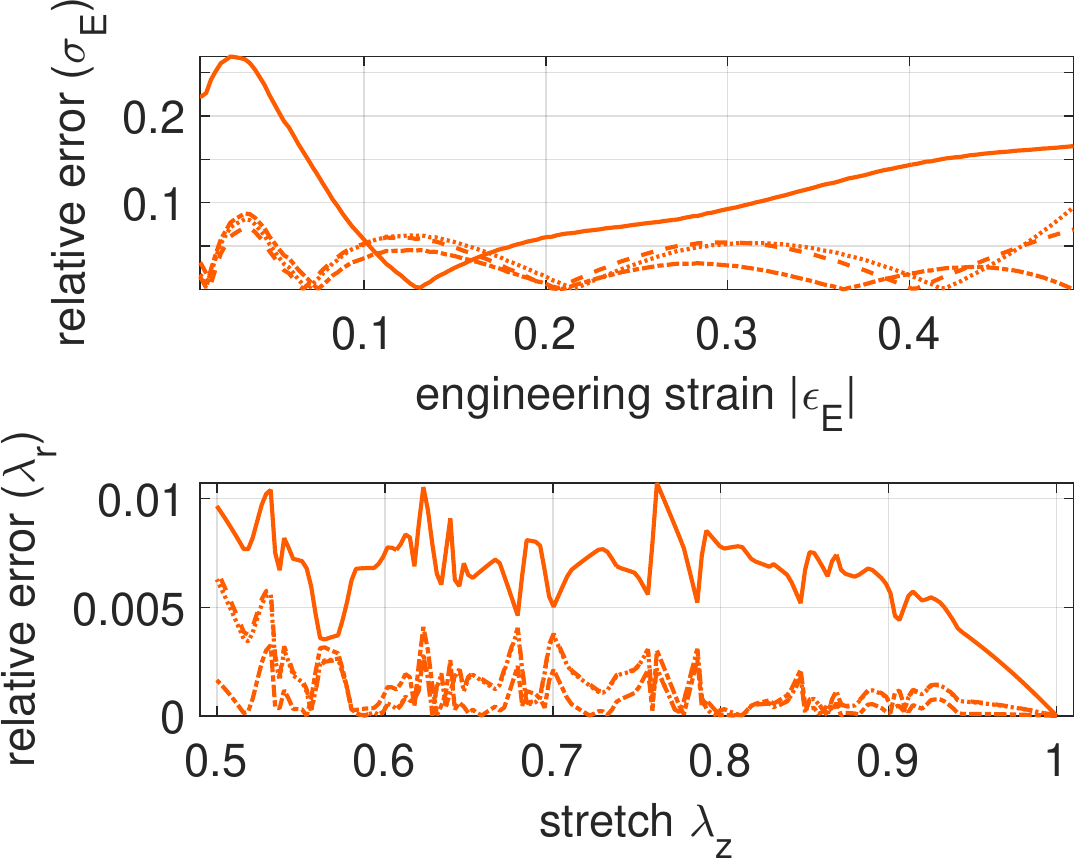}
}
\\
\subfigure[  \label{fig:ogdenII_92040_stress} ]{
\includegraphics[width=0.30\textwidth]{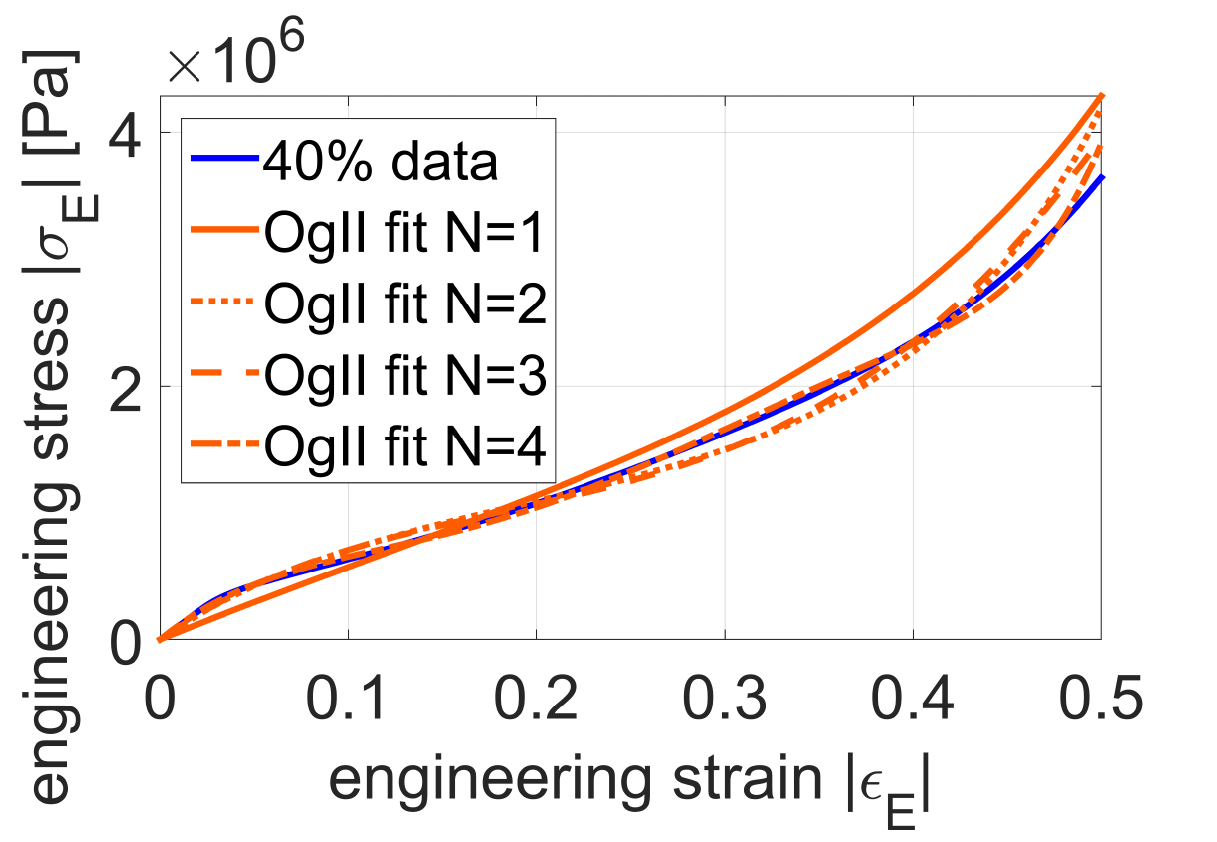}
}
 \subfigure[  \label{fig:ogdenII_92040_tstrain} ]{
\includegraphics[width=0.30\textwidth]{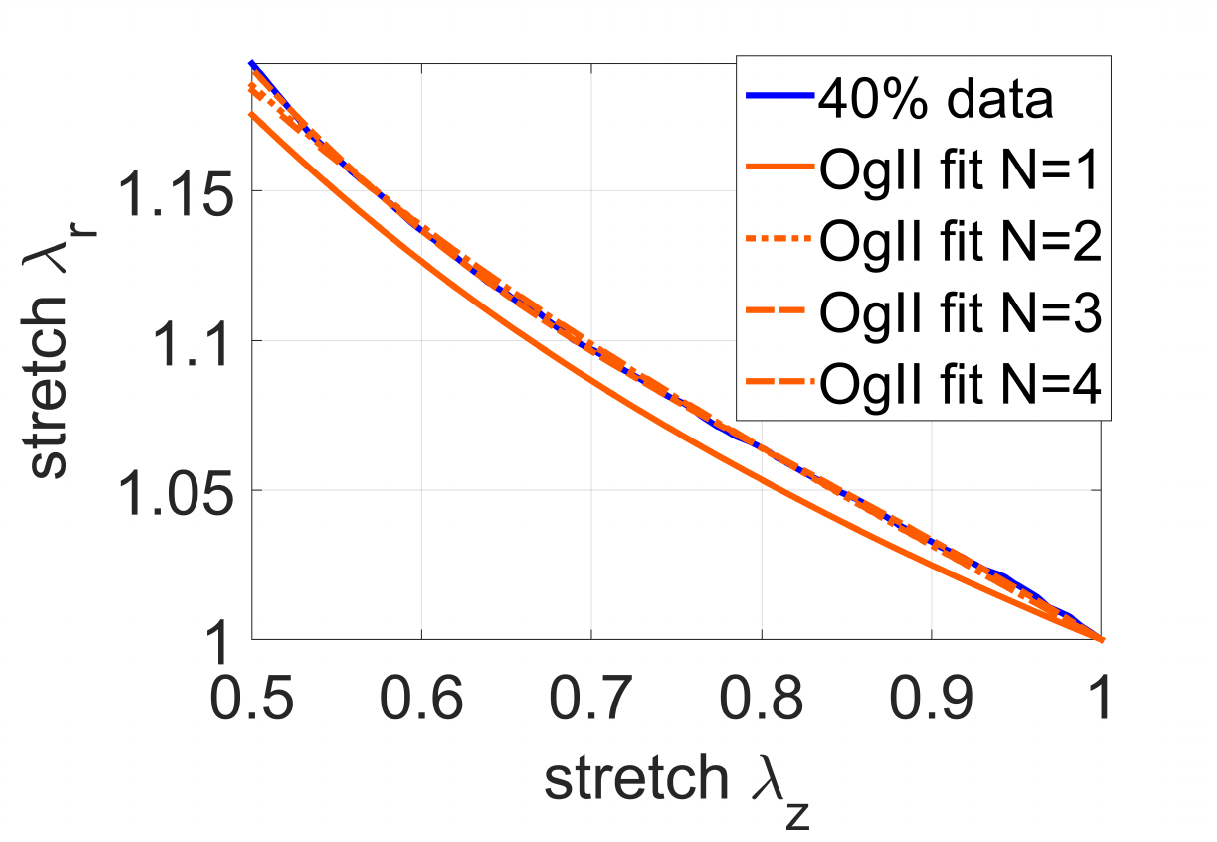}
}
 \subfigure[  \label{fig:ogdenII_92040_relerror} ]{
\includegraphics[width=0.28\textwidth]{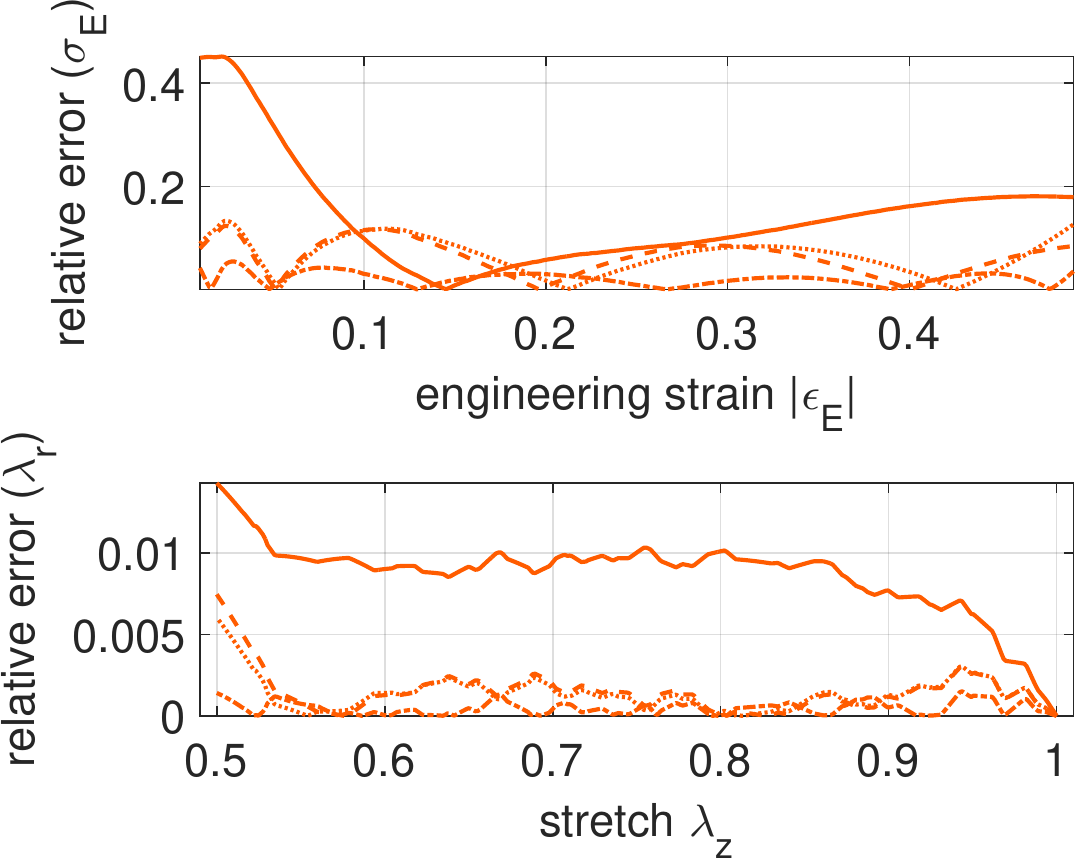}
}

\caption[Comparison of  Ogden Type II  fitted models   against test data for larger filling fractions.]{(Caption in list of figures)   \label{fig:ogdenII_920higher}}
  \end{figure}
\subsection{Compressible Ogden type-II}
 In this model, the strain energy ansatz takes the form
        \begin{subequations}
  \begin{equation}
  \label{eq:ogden}
  W = \sum_{k = 1}^{N} \frac{2 \mu_k}{\alpha_k^2} \left[ \lambda_1^{\alpha_k} + \lambda_2^{\alpha_k}  + \lambda_3^{\alpha_k}  + \frac{1}{\beta_k} \left( J^{-\alpha_k\beta_k} - 1\right)\right],
  \end{equation}
  where $\alpha_k$, $\beta_k$, and $\mu_k$ are   real-valued constants. Under uniaxial compression      
\begin{equation}
\label{eq:sigmaOg}
\sigma_\mathrm{E} = \frac{\partial W}{\partial \lambda_\rmz} = \frac{2}{\lambda_\rmz}\sum_{k = 1}^{N} \frac{  \mu_k}{\alpha_k } \left[ \lambda_\rmz^{\alpha_k} - J^{-\alpha_k\beta_k} \right],
\end{equation}
where $J = \lambda_\rmr^2 \lambda_\rmz$ is obtained by solving the   compressibility relation  
\begin{equation}
\label{eq:sigmaOgCompr}
\sum_{k = 1}^{N} \frac{2 \mu_k}{\alpha_k} \left[ \lambda_\rmr^{\alpha_k} - (\lambda_\rmr^2 \lambda_\rmz)^{-\alpha_k \beta_k} \right] = 0.
\end{equation}
\end{subequations}
 Fig.~\ref{fig:ogdenII_920higher}   presents results for a Compressible Ogden type-II (OgII) strain energy model   at   $\phi = 20\%, \, 30\%,$ and $40\%$ filling fractions (lower filling fractions were not examined with the OgII model in order to avoid the numerical instability issues   encountered by modelling weakly compressible syntactic foams with highly compressible $f(J)$ ansatzes). For high truncation values $N$,   an accurate  description of both the axial stress-strain and transverse strain behaviours was recovered.  Consequently,    OgII strain energy ansatzes  were found to be appropriate forms for   describing the compressive performance of high filling fraction HTM syntactic foams. More generally, this identifies the need for advanced models to describe the response of HTM syntactic foams at large filling fractions. In Table \ref{tab:bestfits} we find that relerr$_\sigma$ and relerr$_\lambda$ for the best-fit Ogden type-II models (corresponding to large $\phi$ HTM syntactic foams) are to within $10.7\%$ at the very worst, with extremely good accuracy recovered in the transverse response. A future approach to modelling these foams could be to construct a SEF with an $f(J)$ motivated by the Hencky models outlined in the previous section as such an investigation lies outside the current scope of the present study.

\subsection{Accuracy of fitted models at small strains} 
\label{sec:lssfsec}
Using   the     coefficients for the best-fit  models described in Table \ref{tab:bestfits} below,    we calculated the associated   small-strain material  constants       to examine the goodness-of-fit to the elastic constants at small strains described in Table \ref{tab:smstrconsts} (full details for all models are included in  the Supplementary Material). We find that the Young's modulus for the $\phi = $0\%, 10\%, 20\%, 30\%, and 40\% samples is recovered to within 2.54\%, 0.53\%, 3.58\%, 1.53\%, and 11.83\% of the measured values respectively.  Interestingly, results for the Ogden type-II ($N=3$) model for $\phi=40\%$ return  more accurate values in this limit, although this model is less accurate over the full strain range. For the Poisson  ratio we find that values are returned with a relative error of $0\%$ for $\phi = $  0\%, 10\%, 20\% samples, and to within 5.41\% and 2.78\% for the $\phi = 30\%$ and $40\%$ samples, respectively.

 \begin{figure}[t]
\centering
\subfigure[  \label{fig:extend_70_unfilled} ]{
\includegraphics[width=0.3\textwidth]{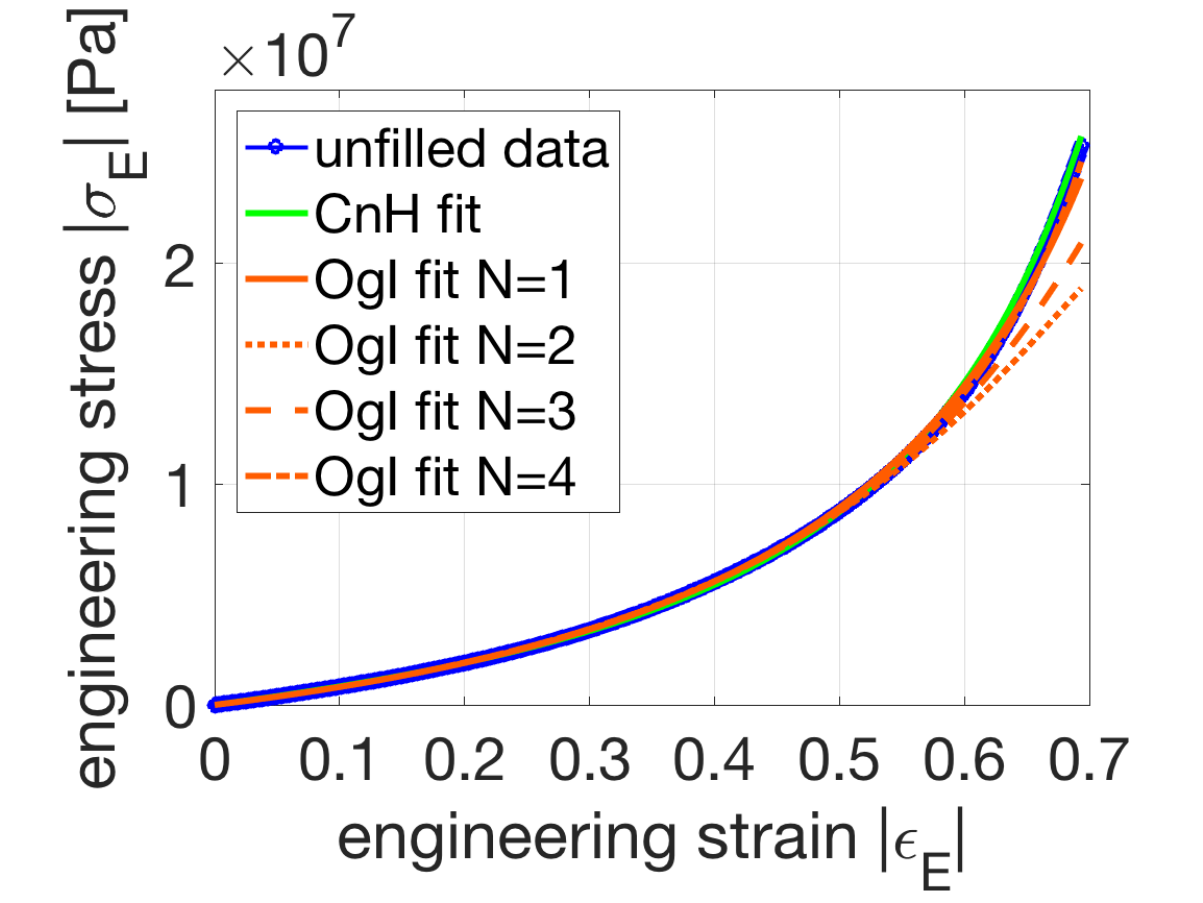}
}
 \subfigure[  \label{fig:extend_70_92040_og1} ]{
\includegraphics[width=0.3\textwidth]{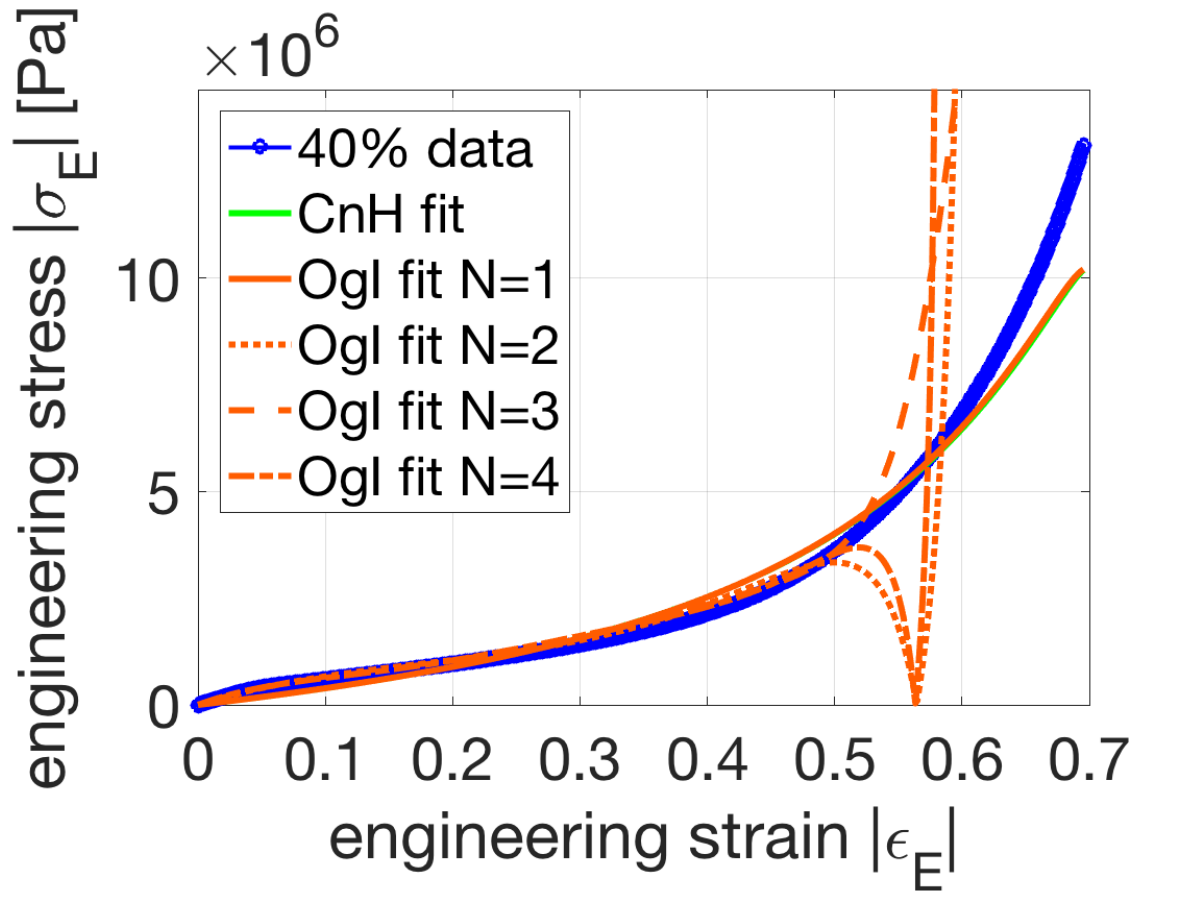}
}
 \subfigure[  \label{fig:extend_70_92040_og2} ]{
\includegraphics[width=0.3\textwidth]{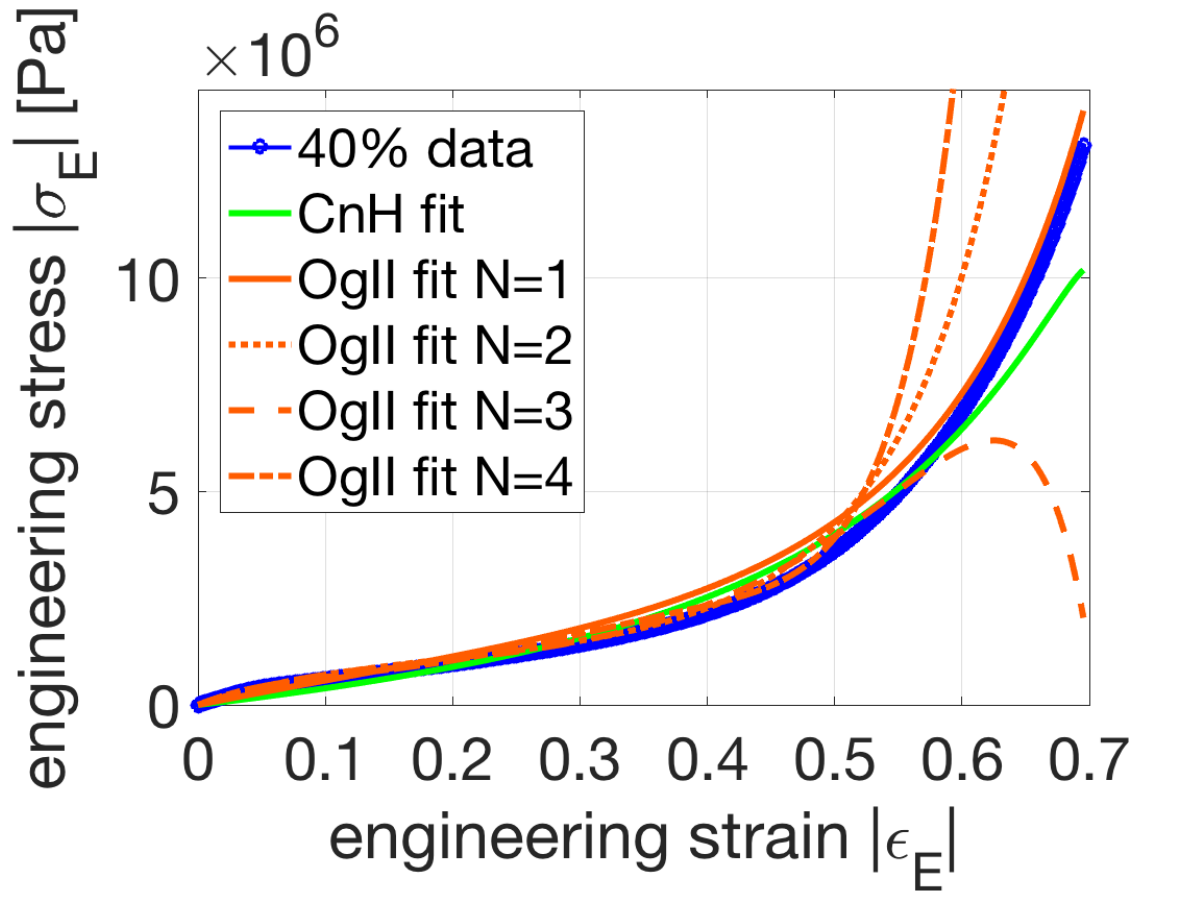}
}
\\
\subfigure[  \label{fig:extend_70_unfilled_relerror} ]{
\includegraphics[width=0.28\textwidth]{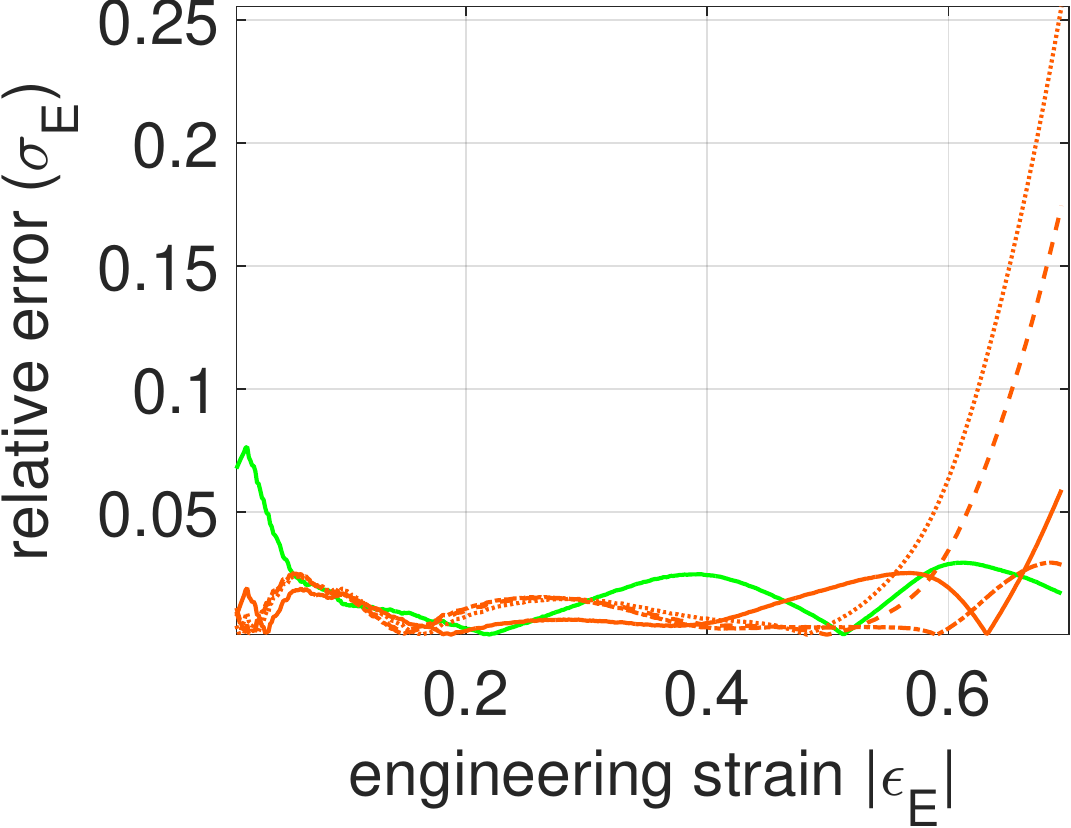}
}
\hspace{3mm}
\subfigure[  \label{fig:extend_70_92040_og1_relerror} ]{
\includegraphics[width=0.28\textwidth]{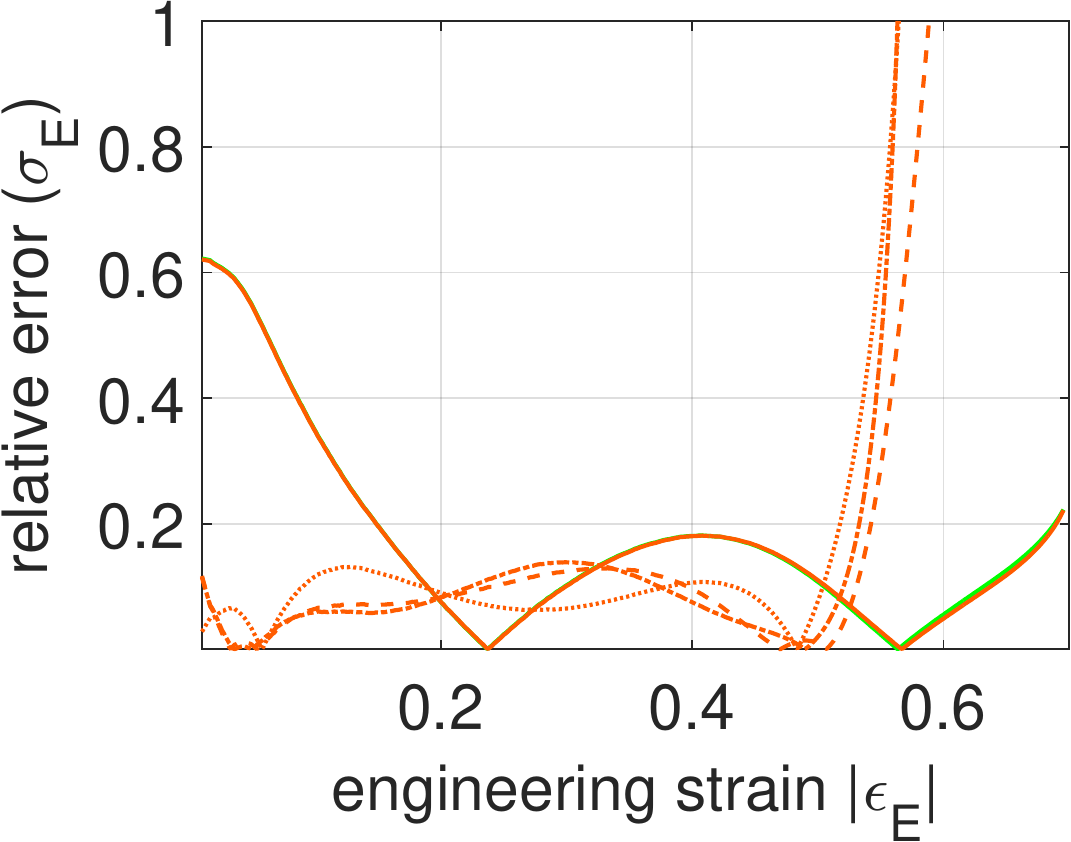}
}
\hspace{3mm}
\subfigure[  \label{fig:extend_70_92040_og2_relerror} ]{
\includegraphics[width=0.28\textwidth]{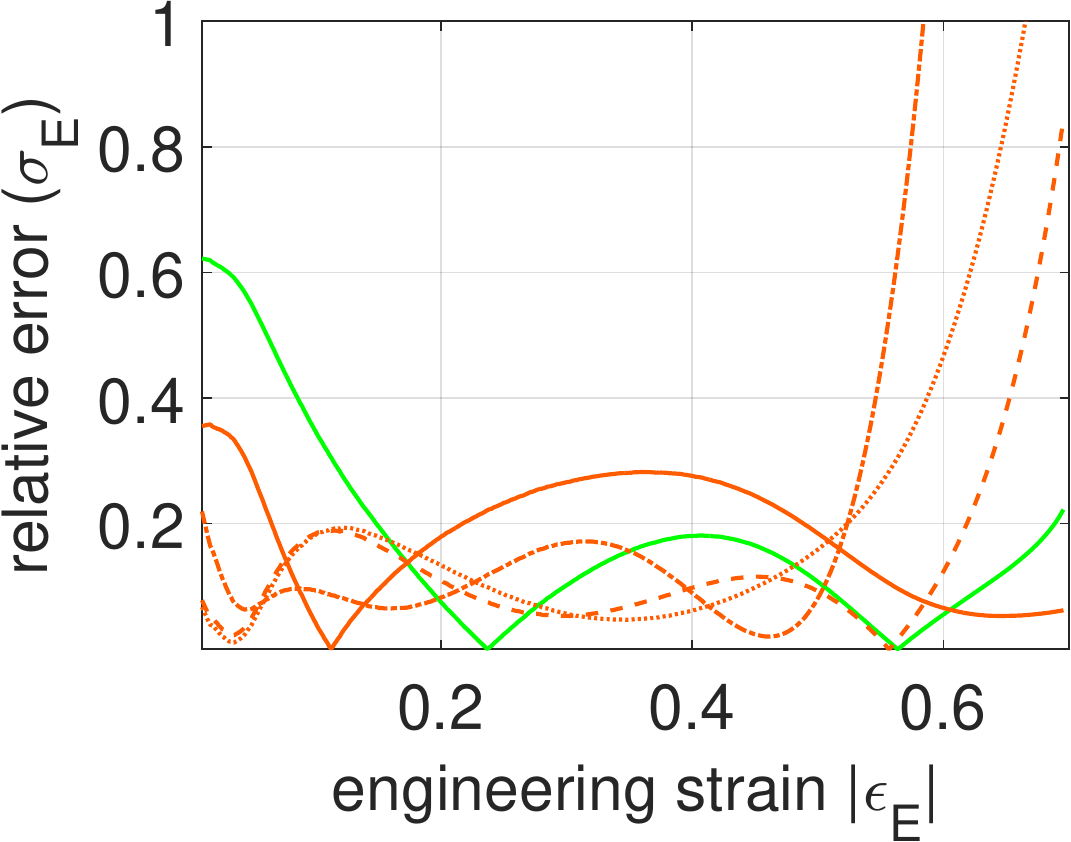}
}
\caption[Test data up to 70\%   strain and   fitted model results   for \subref{fig:extend_70_unfilled} unfilled;    for $\phi=40\%$ foam  with \subref{fig:extend_70_92040_og1}    CnH and Ogden Type I and  \subref{fig:extend_70_92040_og2}    CnH and Ogden Type II models ]{(Caption in list of figures)     \label{fig:extend1}}

  \end{figure}
\subsection{Accuracy of fitted models to even larger strains} 
To test the robustness of our fitted models, we experimentally determined the   stress-strain response of unfilled and 40\% filling fraction foams   to a higher strain level of 70\%, and compared these results against those obtained using the coefficients of the fitted models in previous sections.    Fig.~\ref{fig:extend_70_unfilled}   presents results for $\phi=0\%$,  showing that all fitted models were stable to 70\% strain, with CnH and OgI ($N=1$) being the most accurate.  In the Supplementary Material, maximum relative errors over the range are tabulated, showing that OgI ($N=1$) has max(relerr$_\sigma$) =2.9\% which is much better than the best-fit OgI ($N=3$) model which has max(relerr$_\sigma$) =17.4\%, and highlights the danger of extrapolating results for strain-energy models \cite{ogden2004fitting}.

  For $\phi = 40\%$,  Fig.~\ref{fig:extend_70_92040_og1}  reveals   that all CnH and OgI models were either inaccurate or unstable at higher strains, emphasising the danger of extrapolation  once more, with  Fig.~\ref{fig:extend_70_92040_og2} showing that only OgII ($N=1$) is able to qualitatively describe the response   in the extended strain regime, but with a large quantitative error of max(relerr$_\sigma$) =35.8\%. Such   results motivate  the development of more subtle modelling procedures, such as microstructural models \cite{de2013predicting}. Note that for unfilled polyurethane, these models are extended beyond the elastic strain  regime and into the  plastic deformation region, whereas for $\phi=$ 40\% specimens we did not observe any damage to the samples at strain levels of 70\% (see   \citet{yousaf2020compression}).  
 
 \section{Thermal properties}
 
\subsection{Testing Methods}
Differential scanning calorimetry (DSC) testing was performed by employing a Q100 DSC (TA Instruments) using aluminium hermetic pans. The weight of each sample  was $\sim$6 mg, and three samples were tested for each foam configuration. The tests were conducted at a ramp rate of 10$^\circ$C/min from -90$^\circ$C to 300$^\circ$C under a constant flow of nitrogen (50 mL/min)

\begin{figure}[t]
\centering
\subfigure[  \label{fig:heatflowall} ]{
\includegraphics[width=0.45\textwidth]{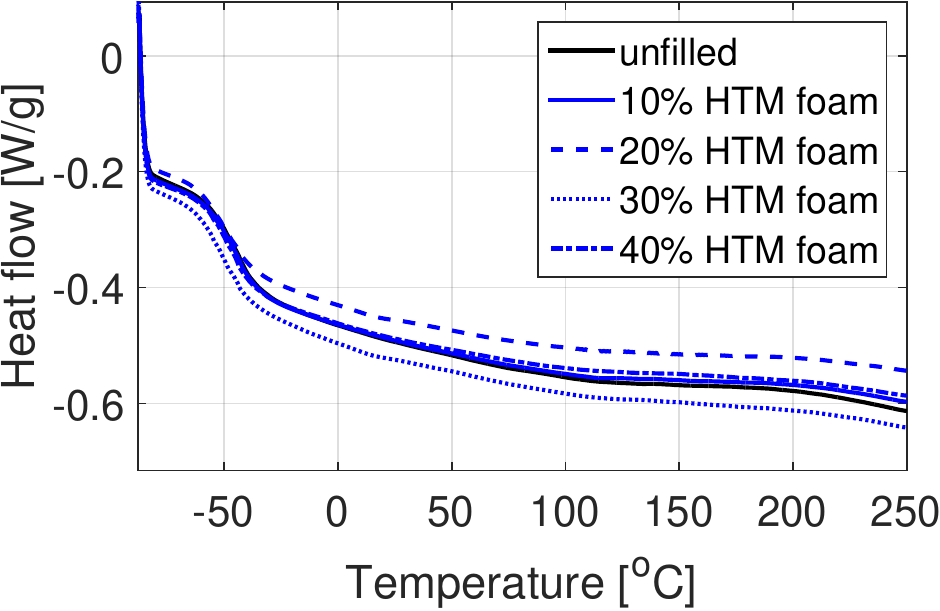}
}
 \subfigure[  \label{fig:heatflowexp} ]{
\includegraphics[width=0.45\textwidth]{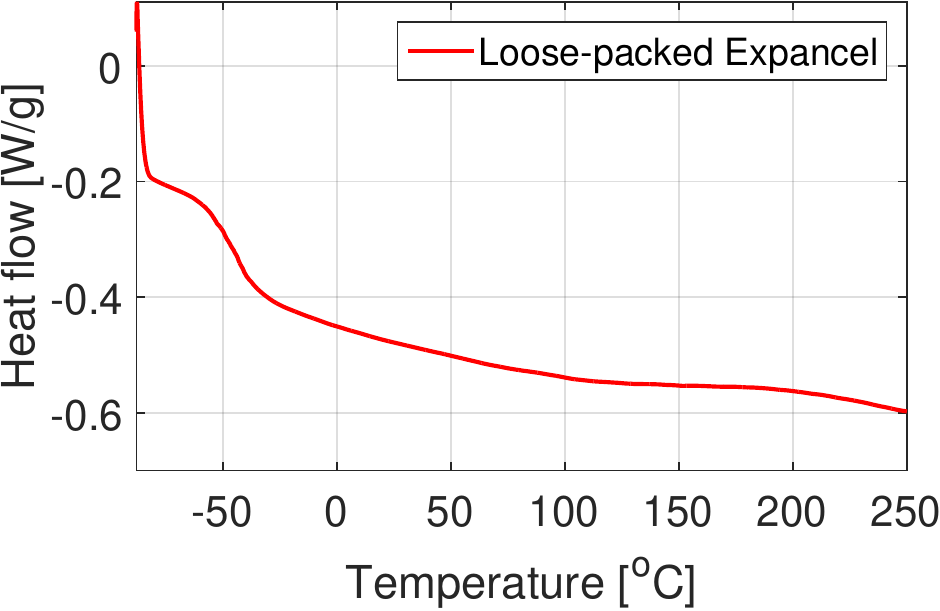}
}
\caption[Test data for heat flow obtained by DSC for \subref{fig:heatflowall} all syntactic foam samples  and    \subref{fig:heatflowexp} a loosely-packed Expancel sample]{(Caption in list of figures)     \label{fig:figthermal}}

  \end{figure}
\subsection{Experimental results}
Fig.~\ref{fig:figthermal} shows the heat flow (per gram of material) against temperature for all foam samples as well as for a  hermetic pan loosely-packed with Expancel microballoons.   Fig.~\ref{fig:heatflowall}   shows that the presence of Expancel microspheres has a minimal impact on the thermal performance of the matrix. The glass transition temperature $T_g$ is also largely unaffected by changes in filling fraction ($T_g \approx -50^\circ$C), which is consistent with  earlier   literature \cite[Table 5-15]{lu1999processing}. In Fig.~\ref{fig:heatflowexp} we present   results for the loosely-packed Expancel microballoons and find a qualitatively similar heat flow profile to the polyurethane matrix in Fig.~\ref{fig:heatflowall}, which explains the minimal change observed in thermal properties of HTM syntactic foams at all concentrations. For reference,  polyurethane  matrix syntactic foams containing HGMs have been reported to exhibit glass transitions $T_g> 100^\circ$C suggesting that HGMs  exhibit a significantly different heat flow profile \cite{otloski1990polyurethane}.

\section{Conclusions} \label{sec:concl}
An investigation to characterise the   compressive performance  of HTM syntactic foams   has revealed increased small-strain stiffness with increasing filling fraction $\phi$, strong stiffness recoverability, and a non-trivial transverse stretch relationship for large $\phi$. Strain energy models were constructed to describe both the stress-strain and transverse strain response of the foams under uniaxial compression, where it was found that  OgI  models achieve the best fits   for   filling fraction    foam  up to $\phi = 20\%$, as they recover accurate small-strain constants,     accurate qualitative descriptions up to 50\% strain, and possessed the smallest relative errors up to strains of 50\%. For higher filling fractions,  OgII   strain energy models were able to describe HTM syntactic foams, up to strains of 50\%,  due to the more general form of the compressibility condition $f(J)$. Our findings highlight a major shortfall in compressible strain energy models, in particular the lack of stability observed when extending fitted models to 70\% compressive strain, which motivates the development of more general compressible strain energy ansatzes $W$, as well as micromechanical models, for characterising foams in the future. Nonetheless, the strain energy descriptions   obtained       are anticipated to prove   useful   for industrial applications and    for the      validation of future micromechanical models. The stiffness recoverability properties of these foams  show considerable promise when coupled with their  small-strain stiffening response. We also find that the thermal characteristics of our polyurethane matrix are generally unaffected by the presence of HTMs, which provides an additional motivation for  the use of HTMs as a lightweight filler for thermal applications. Finally, this work   provides an important foundation for studies into the time-dependent (viscoelastic) behaviour of HTM syntactic foams, building on results from cyclic testing which showed that HTMs had an effect on the dissipative loss \cite{yousaf2020compression}, as well as for the digital design of materials with optimised thermal and mechanical properties.

\section*{Acknowledgments}
Alison Daniel (Thales UK) for sample manufacture.  Chloe Loveless and Damindi Jones (Department of
Materials, University of Manchester) for their assistance with DSC experiments.

\section*{Funding Statement}
This work was supported by the EPSRC (grant numbers EP/L018039/1 and
 EP/R014604/1) and Thales (Thales Research  Structures, Materials and Acoustics Research Technologies (SMART) Hub).




\bibliographystyle{elsarticle-num-names}

 
 \bibliography{papers}

\listoffigures

\clearpage
\listoftables

\section*{Tables}
\clearpage
\begin{table}[t]
\centering
\begin{tabular}{lccccc}
							 &  Particle diameter (Median)  & Shell thickness  &  Bulk density   \\ \hline
Manufacturer-supplied \cite{AkzoNobel2016techsheet,Alberts2018correspondence} & 55-85 $\mu$m & 0.35 $\mu$m  &   $30\pm 3$ kg/m$^3$ \\
Experimentally-measured \cite{curd2020characterisation,daniel2017correspondence} & 83.8 $\mu$m & 0.29 $\mu$m  &   28.7  kg/m$^3$ 
\\ \hline
\end{tabular}

\caption[ Properties of    Expancel 920 DE hollow thermoplastic   microspheres ]{(Caption in list of Tables) \label{tab:table1}}

\end{table}

\begin{table}[t]

\centering
\begin{tabular}{rrcccccc}
  &   &   $Y$ \,$\left[ \mathrm{MPa}\right]$&   $\nu$\,  &   $\rho \, \left[ \mathrm{kg}/\mathrm{m}^3\right]$  & $Y/\rho \, \left[ \times 10^3 \, \mathrm{m}^2/\mathrm{s}^2\right]$    \\ \hline \hline
  & 0\%   &   7.08& 0.49 &1083  & $6.5  $  \\  
&10\%    &  7.53 & 0.46& 981 & $7.7   $     \\
  &20\%    & 8.09  & 0.44 &881 & $9.2 $    \\
    &30\%    &8.52  & 0.37 &779  & $11  $    \\
 &40\%    & 10.9 & 0.36 &672 & $16.3  $   \\ \hline \hline
\end{tabular}

\caption[Test values for small-strain  bulk properties of HTM syntactic foams including Young's modulus $Y$,    Poisson's ratio $\nu$,   density $\rho$, and specific stiffness $Y/\rho$.]{(Caption in list of Tables)
 \label{tab:smstrconsts}}

\end{table}

\begin{table}[t]
\centering
\begin{tabular}{lccccc}
 & Model & N& Max. rel. error $\sigma$ 	 &		Max. rel. error $\lambda_\mathrm{r}$ \\ \hline
0\% & OgI &3 & 0.020845 & 0.010557 \\
10\% & OgI &   4 & 0.042669 & 0.0067259  \\ 
20\% & OgI &   4 & 0.086749 & 0.059044  \\ 
30\% & OgII & 3 & 0.079871 & 0.0067287  \\ 
40\% & OgII &    4 & 0.10736 & 0.0015126  \\ 
  \hline
\end{tabular}

\caption[Summary of maximum relative errors over entire strain range for best-fitted models    ]{(Caption in list of Tables) \label{tab:bestfits}}

\end{table}

\end{document}